\documentclass[11pt,a4paper,english,twoside]{article}

\usepackage{a4wide}
\usepackage{amssymb, amsmath}
\usepackage{graphicx}
\usepackage[all]{xy}
\usepackage{enumerate}
\usepackage{dsfont}
\usepackage{empheq}
\usepackage{cite}
\usepackage{float}
\usepackage{wasysym}
\usepackage{makecell}
\usepackage{colortbl}

\usepackage{subcaption}

\usepackage[utf8]{inputenc}
\usepackage[T1]{fontenc}

\usepackage[pdftex,hyperref,x11names]{xcolor}
\usepackage[pdftex,colorlinks=true,
pdfstartview=FitV,
pdfnewwindow=true,
linktoc = page,
linkcolor= RoyalBlue3,
citecolor= Red3,
urlcolor= RoyalBlue3,
hyperindex=true,
hyperfigures=false]{hyperref}

\newcommand{\beq}{\begin{equation}}
\newcommand{\eeq}{\end{equation}}
\def\bea#1\eea{\begin{align}#1\end{align}}
\def\beal#1\eeal{\begin{subequations}\begin{align}#1\end{align}\end{subequations}}
\newcommand{\nn}{\nonumber}

\newcommand{\va}{\vec{\alpha}}
\newcommand{\vb}{\vec{\beta}}
\newcommand{\vg}{\vec{\gamma}}
\newcommand{\vv}{\vec{\varphi}}
\newcommand{\vx}{\vec{x}}
\newcommand{\vn}{\vec{n}}

\newcommand{\eq}[1]{\begin{equation}#1\end{equation}}
\newcommand{\spl}[1]{\begin{split}#1\end{split}}

\def\d {{\rm d}}

\begin{document}
\numberwithin{equation}{section}

\begin{titlepage}
	\vspace{14pt}
	
	\begin{center}
		
		{\LARGE \bf 
		Universal  Cosmologies 
		 }\\
		
		\vspace{1.5cm}
		
		{\bf \large Paul Marconnet and Dimitrios Tsimpis }
		
		\vspace{0.3cm}
		
		{\it   Institut de Physique des Deux Infinis de Lyon   }\\
		{\it Universit\'e de Lyon, UCBL, UMR 5822, CNRS/IN2P3 }\\
		{\it 4 rue Enrico Fermi, 69622 Villeurbanne Cedex, France  }
		
		\vspace{0.3 cm}
		
		\texttt{ \href{mailto:marconnp@gmail.com}{marconnp@gmail.com}, \quad \href{mailto:tsimpis@ipnl.in2p3.fr}{tsimpis@ipnl.in2p3.fr}}
			
		\vspace{1.5 cm}
		
	\end{center}

\vspace{2cm}
\abstract{\noindent 
Universal cosmologies  are exact  solutions of 10d type IIA supergravity containing  a 4d Friedmann-Lema\^{i}tre-Robertson-Walker  factor, that can   also be repackaged as 
solutions of 4d models, i.e.~as 4d consistent truncations.~We extend  the dynamical system analysis of  universal cosmologies,  beyond 
the case of a single exponential potential.~For an open universe (negative 3d spatial curvature), 
these models generally possess many desirable features:~parametric control of e-folds, late-time acceleration   from potentials with steep exponentials (i.e.~in accordance with swampland bounds), 
small string-loop and  $\alpha'$-corrections, scale separation and/or absence of decompactification.
}

\vfill\break
\end{titlepage}
\tableofcontents

\section{Introduction}

Obtaining realistic four-dimensional  (4d) cosmologies from higher-dimensional supergravities arising from  the low-energy limit of superstring theory is an interesting open question.~Although 
it has been known for some time that  time-dependent compactifications can lead to solutions exhibiting  cosmic acceleration  \cite{Townsend:2003fx}, 
thereby evading the   no-go theorem of \cite{Gibbons1,Gibbons:2003gb,Maldacena:2000mw},\footnote{See  \cite{Russo:2018akp,Russo:2019fnk} for an improved version of the no-go that comprises the case of time-dependent compactifications.} it was believed until recently that whenever such 
models  \cite{Ohta:2003pu,Ohta:2003ie,Ohta:2004wk,Roy:2003nd,Gutperle:2003kc,Emparan:2003gg,Townsend:2003qv} 
exhibit transient acceleration,  the  number of e-folds can be at most order one  \cite{Chen:2003dca,Wohlfarth:2003kw}.

In \cite{Marconnet:2022fmx}  we studied time-dependent solutions of Type IIA 10d supergravity with a 4d 
Friedmann-Lema\^{i}tre-Robertson-Walker  (FLRW) factor, 
compactified on different classes of 6d manifolds: Calabi-Yau, Einstein, Einstein-K\"{a}hler.~The cosmologies  thereby obtained  were called  ``universal'',  in that they only depend on the general features of the compactification manifolds.~In addition we established a cosmological consistent truncation of IIA supergravity to a 4d gravitational theory coupled to two scalar fields,   i.e.~a repackaging of the 10d equations of motion such that every FLRW  solution of the 4d theory lifts to a  solution of 10d IIA supergravity.

Our work  found   that cosmologies — with or without Big Bang singularities — featuring  late-time   acceleration, or rollercoaster \cite{DAmico:2020euu} 
asymptotic behavior, are generic  in flux compactifications of 10d supergravity.~Morever, 
we showed that solutions exhibiting  transient acceleration with a parametric control of e-folds are not only possible, 
but are also generic.~The crucial ingredient in achieving these features  is 
provided by the negative 3d spatial curvature:~all these solutions require an open universe,~i.e.~a FLRW cosmology with hyperbolic 3d spatial slices.~Interestingly, an open universe has been argued to be a generic prediction of the string landscape  \cite{Freivogel:2005vv}.~The recent results of \cite{Bedroya:2025ris} provide an additional compelling 
motivation to consider open cosmologies.

Ref.~\cite{Andriot:2023wvg} put this analysis in the general context of 4d  effective models with an exponential potential, and discussed their realization in string theory.~It was explained therein 
that the steepness of the exponential potentials typically coming from supergravity (therefore in asymptotic regions of string theory, and in accordance with  
the   swampland bounds  \cite{Obied:2018sgi,Hebecker:2018vxz,Andriot:2019wrs,Lust:2019zwm,Bedroya:2019snp,Andriot:2020lea,Rudelius:2021oaz,Rudelius:2021azq}),  in conjunction with the 
negative 3d curvature, 
guarantee the existence of the attractor fixed point on the boundary of the acceleration region —   responsible for the   features of the solutions of   \cite{Marconnet:2022fmx}  mentioned in the previous paragraph.~Moreover, it was emphasized  that all these solutions have vanishing cosmological horizon, and conjectured  that this may be a generic feature
of  theories of  quantum gravity  (see also \cite{Boya:2002mv,Townsend:2003qv}).

Cosmological models have attracted a renewed interest  within string theory  
\cite{Agrawal:2018own, Olguin-Trejo:2018zun, Hebecker:2019csg, Cicoli:2020cfj, ValeixoBento:2020ujr, Cicoli:2021fsd, Rudelius:2022gbz, Calderon-Infante:2022nxb, Shiu:2023nph, Shiu:2023fhb, 
Cremonini:2023suw,  Hebecker:2023qke, Freigang:2023ogu, 
VanRiet:2023cca, Shiu:2023yzt, 
Andriot:2024jsh, Shiu:2024sbe, Casas:2024oak, 
 Andriot:2024sif,   Andriot:2025cyi},  
not least because of recent   data providing evidence for a dynamical dark energy  \cite{DESI:2025zgx, DESI:2025fii}.~Although steep 
single exponential models   appear to be excluded by current observations \cite{Bhattacharya:2024hep, Alestas:2024gxe,Akrami:2025zlb},  their  multi-field, multi-exponential, spatially curved extensions warrant further study.~In the present paper, we extend  the dynamical system analysis of the  universal cosmologies of  \cite{Marconnet:2022fmx} beyond 
the case of a single exponential potential, with a particular focus on  the late-time physical properties   of these models.

Our analysis confirms the presence of several phenomenologically desirable features.~In addition to the aforementioned properties already noted in  \cite{Marconnet:2022fmx}, we find that  both 
loop corrections in the string coupling ($g_s$-corrections) and higher-order derivative corrections  ($\alpha'$-corrections) 
are generally suppressed at late times.~Moreover, several models exhibit scale separation at late times, i.e.~the ratio of the effective radius of their internal space to the 4d Hubble length  
vanishes at late times. This guarantees that, in addition to being 10d supergravity solutions,  
these  models are also bona fide 4d effective descriptions.

The question  of scale separation in cosmological solutions was recently revisited in \cite{Andriot:2025cyi}, where it was noted that cosmologies supported by the curvature of the internal compactification space do not exhibit late time scale separation.~We show that models in which the 6d curvature contributes non-trivially to the 4d potential at late times — even if its contribution is not dominant — also fail to exhibit scale separation.~Instead, these models show an absence of decompactification, i.e.~the effective radius of the internal space scales as the 4d Hubble length at late times.~This generalizes  the results of \cite{Andriot:2025cyi}, which focused on curvature-dominated potentials, to the cases where the potential merely includes a curvature contribution, in line with the results of \cite{Shiu:2023fhb} for flat universe.

In Section~\ref{sec:2} we introduce  the general framework of our models, from the point of view of a 4d gravitational theory minimally coupled to two scalar fields.~At this stage we remain agnostic as to the  possible  higher-dimensional origin of this theory, aiming to provide the most general analysis.~We introduce  the dynamical system description of the part of the phase space that is independent of the details of the potential of the theory, the possible critical points and corresponding analytic solutions.

In Section~\ref{sec:single} we specialize to the case of a single exponential potential.~We review the critical points and their stability properties.~In 
Section~\ref{sec:contrunc} we recall  that in this case  there  always exists   a consistent sub-truncation to a single scalar field, resulting 
in a theory whose potential generally has a steeper exponential than the original theory before truncation.~In Section~\ref{sec:heteroc} we  give the full analytic form of a cosmological  solution exhibiting eternal acceleration and no Big Bang singularity.

In Section~\ref{sec:2exp} we move on to the case of a two-exponential potential.~We employ a set  of  phase space variables —  different from  the ones typically used in dynamical system descriptions of multi-exponential systems —  that offer 
a visual, intuitive description of motion in  phase space. The latter now contains a two-dimensional sub-plane, which can be thought of as the space of exponents of the potential.~Points  in that plane correspond to an  effective exponent,   
 constrained to move on a certain straight  line.~Moreover, the effective exponent at the stable critical point is given by  the shortest vector connecting the origin of the plane to the convex hull of the two exponents of the potential.

 In Section~\ref{sec:universal}, we apply the tools developed above to the universal-cosmology models of  \cite{Marconnet:2022fmx}, with a particular focus on their late-time behavior. We find that, in all but one model, both higher-order $g_s$- and   $\alpha'$-corrections 
 are suppressed at late times.~Moreover, all but one model avoid  decompactification at late times, and three of these additionally exhibit scale separation. 
 Finally, in Section~\ref{sec:multi} we comment on the multi-exponential case.~We conclude in Section~\ref{sec:conslusions} with a discussion of future directions.

 \section{Two-scalar model}\label{sec:2}

Our starting point is the 
 4d Einstein action with two minimally-coupled scalar fields,
\eq{\label{2sc}
S_{4\d}=\int\d^4 x\sqrt{-g}\left(
\frac{1}{16\pi G}R-\frac12 g^{\mu\nu} \sum_{i=1}^2 \partial_\mu \varphi_i\partial_\nu \varphi_i
  -V(\varphi_1,  \varphi_2)
\right)~,}
where $G$ is the 4d Newton's constant.  In Section~\ref{sec:universal} we will interpret $\varphi_{1,2}$ as originating from the dilaton and the warp factor  in 10d supergravity, but for now 
 we will remain agnostic as to the potential higher-dimensional embedding of the action \eqref{2sc}.

We are interested in cosmological solutions of     FLRW  form,
\eq{\label{metric4}
\d s^2 = - \d t^2 + a(t)^2 \left(\frac{\d r^2}{1 -k r^2} + r^2 \d \Omega^2 \right)~;~~~a(t)>0 
~,}
where $a(t)$ is the scale factor and $k=\pm1$, $k=0$  corresponds to a closed, open, flat  4d universe respectively. 
Assuming homogeneous scalar fields, the   matter equations of motion read, 
\eq{\label{2bgfieldeom}
\ddot{\varphi_i}
+3H\dot{\varphi_i}+\partial_{\varphi_i}V(\varphi)=0 ~;~~~i=1,2
~,}
where $H:=\frac{\dot{a}}{a}$ is the Hubble parameter, and a dot stands for derivative with respect to the cosmological time $t$.  
The gravitational equations of motion are given by 
  the two  Friedman equations,\footnote{All  equations of motion are invariant under the rescaling $k\rightarrow a^2_0 k$, $a\rightarrow a_0 a$, for any constant $a_0>0$.}
 \eq{\spl{\label{2bgF}
H^2&=\frac{8\pi G}{3}\rho-\frac{k}{a^2}\\
\dot{H}&=-4\pi G \sum_{i=1}^2\dot{\varphi_i}^2+\frac{k}{a^2}
~,}}
where the energy density and pressure are given by, 
\eq{\label{2set}
\rho=\frac12 \sum_{i=1}^2\dot{\varphi_i}^2+V(\varphi)~;~~~p=\frac12  \sum_{i=1}^2\dot{\varphi_i}^2-V(\varphi)
~.}
Moreover, it can be seen that, together with Eqs.~\eqref{2bgfieldeom}, the first Friedman equation implies the second one.

\subsection{Dynamical system}

In order to rewrite the equations of motion in the form of a dynamical system, let us define  \cite{Copeland:1997et}, 
\eq{\label{26}
\gamma_i (\varphi):= -\partial_{\varphi_i}\ln V ~;~~~i=1,2
~,}
which are not constant in general. 
Furthermore,  we   set $8\pi G=1$ and define the  variables,\footnote{\label{f1}A negative potential can also be accommodated,  by  defining 
$z = \frac{\sqrt{|V|}}{H \sqrt{3}}$, which results in flipping the sign of $z^2$ in \eqref{system2},\eqref{1stFried2}.  This implies in particular that there can be no acceleration if the  potential is negative, as can be seen 
by  flipping the sign of $z^2$ in \eqref{210}. For this reason we 
restrict to non-negative potentials.}
\beq
N:= \ln a \ ,\quad x_1 := \frac{\dot{\varphi_1}}{H \sqrt{6}} \ ,\quad 
x_2 := \frac{\dot{\varphi_2}}{H \sqrt{6}} \ ,\quad
z := \frac{\sqrt{V}}{H \sqrt{3}} \ ,\qquad H\neq 0 \ ,\ V\geq0 \ .\label{variables2}
\eeq
In the following it will be useful to introduce a two-component vector notation, such that $\vec{x}=(x_1,x_2)$, $\vg=(\gamma_1,\gamma_2)$, etc. 
We thus obtain the following system of equations,
\eq{\spl{ \label{system2}
& \vec{x}^{~\!\prime}=  \sqrt{\frac{3}{2}}\, \vg \, z^2   + \vec{x} \, \Big( 2\vec{x}^{~\!2} - z^2-2 \Big) \\
&  z'= z \left( -\sqrt{\frac{3}{2}}\, \vg\cdot\vec{x}  +  2\vec{x}^{~\!2} -z^2+1 \right)  \ ,
}}
where a prime denotes derivative with respect to $N$, 
together with the constraint,
\eq{
\vec{x}^{~\!2} +z^2 =  1+ \frac{k}{\dot{a}^2} ~.\label{1stFried2}
}
Eq.~\eqref{1stFried2} only needs to be imposed at some initial time, as it is consistently propagated by the remaining equations.~Eqs.~\eqref{system2}, supplemented by the constraint \eqref{1stFried2}, are 
a rewriting of 
\eqref{2bgfieldeom}, \eqref{2bgF} in terms of the new variables.~However \eqref{system2} is not an autonomous dynamical system in general, unless $\vg$ happens to be constant.

Even without knowledge of the dynamics of $\vg(\varphi)$, there are a few important conclusions that we can draw from this system. 
Firstly, from the definition of the variable $z$, it follows that expanding cosmologies correspond to trajectories with $z>0$. 
Secondly, from \eqref{2bgF}, \eqref{variables2} we have, 
\eq{\label{210}
\frac{\ddot{a}}{a}=H^2(z^2-2\vx^{~\!2})
~,}
from which we see that  accelerating cosmologies correspond to trajectories inside the 
 cone  in $(\vx,z)$-space  given by,
\eq{\label{52}
\mathcal{A}=\left\{(\vx,z)\in\mathbb{R}^3~|~z^2>2\vx^{~\!2}\right\}
~.}
Thirdly,  it follows from   \eqref{system2} that, 
\eq{\label{54}
\frac12\left(\vec{x}^{~\!2} +z^2\right)'
=
(z^2-2 \vec{x}^{~\!2} ) (1-\vx^{~\!2} -z^2)
~.}
Therefore the unit  sphere,
\eq{\label{ginvs}
\mathcal{S}=\left\{(\vx,z)\in\mathbb{R}^3~|~\vx^{~\!2}+z^2=1\right\}
~,}
is an invariant surface.  This means that trajectories in the $(\vx,z)$-space that cross $\mathcal{S}$ at one point, must be entirely contained in $\mathcal{S}$. Furthermore, 
trajectories that have one  point in the interior  (exterior) of   $\mathcal{S}$,  must lie entirely in the interior (exterior) of $\mathcal{S}$. 
It then follows from \eqref{1stFried2} that open (closed) FLRW cosmologies correspond to trajectories in the interior  (exterior) of $\mathcal{S}$, while flat cosmologies correspond to 
trajectories on $\mathcal{S}$.

In the following we will  restrict our attention to FLRW cosmologies with $k\leq0$.\footnote{The reason for this restriction is that, for the systems we are examining here, there can be no stable critical points with $k>0$ \cite{Hartong:2006rt, Andriot:2023wvg}. In particular, closed FLRW universes cannot 
exhibit (semi-)eternal acceleration.} We can then see that, as a 
consequence of \eqref{52},\eqref{54},  trajectories inside (outside) the acceleration region $\mathcal{A}$ always move towards increasing (decreasing)  distance from the origin of the $(\vx,z)$-space.

\subsection{Critical points}\label{sec:critical}

\begin{figure}[H]
\begin{center}
\includegraphics[width=.5\textwidth]{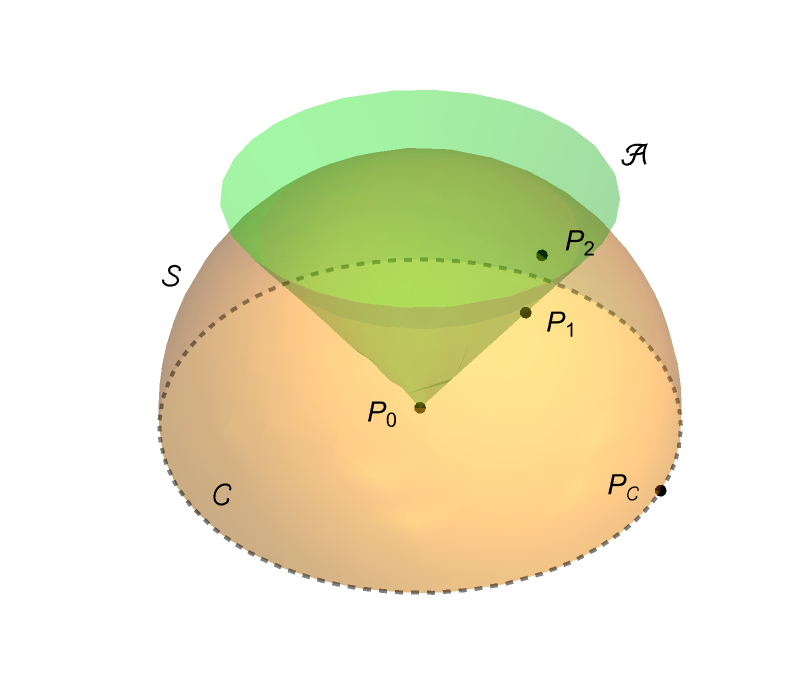}
\caption{\!Critical points of the dynamical system, Eqs.~\eqref{system2}, \eqref{1stFried2}.~The horizontal plane, vertical axis  are parameterized by the coordinates $\vx$, $z$ respectively.~The northern hemisphere of the unit surface $\mathcal{S}$ is depicted — whose interior corresponds to expanding, open cosmologies.~The  green  cone is the acceleration region $\mathcal{A}$.~We depict  a critical  point $P_{\mathcal{C}}$ on the equator, i.e.~the   circle $\mathcal{C}$  of kinetic energy  dominated  fixed points (dashed  line); the curvature dominated critical point $P_0$ at the origin; the  critical  point $P_2$ with  
$\gamma^2=1.4$ on $\mathcal{S}$, and in the interior \!of \!$\mathcal{A}$; the critical point $P_1$ with $\gamma^2=3.6$  on the boundary of \!$\mathcal{A}$, \!and  in the interior of $\mathcal{S}$.}\label{fig_2a1}
\end{center}
\end{figure}

Independently of the dynamics of $\vg(\varphi)$,  critical points of the  system must necessarily be critical points of \eqref{system2}, i.e.~solutions of $\vx^{~\!'}=0=z'$ for $\vg=\vg_*$, where the constant vector $\vg_*$ is  the value of $\vg$ at the critical point.~In addition, the solution must satisfy the constraint \eqref{1stFried2}.  This  leads to the following list of potential   critical points, also depicted in Figure~\ref{fig_2a1}:

$\bullet$  All points  $P_{\mathcal{C}}
\in 
\mathcal{C}$ of the unit circle at the equator, 
\eq{
\mathcal{C}:= \left\{(\vx,z)\in\mathbb{R}^3~|~
\vx=\vec{n}~\text{and}~z=0\right\}
~,}
where $\vec{n}$ is a constant   unit vector,  $|\vec{n}|=1$.~Being   on the unit sphere $\mathcal{S}$, they require $k=0$. Moreover the condition $z=0$ implies $V=0$.  These points  correspond to {\it kinetic domination}.

\vfill\break


$\bullet$ The origin $P_0=(0,0,0)$ of the $(\vx,z)$-space.
This point is in the interior of  $\mathcal{S}$  and has $z=0$,   it thus requires $k=-1$ 
 and $V=0$.~In addition we must have $\dot{a}^2=1$ for the constraint to be satisfied, which implies vanishing acceleration.~This point corresponds to {\it curvature domination}.

$\bullet$ The point $P_1=(\vx,z)$ with  $z^2=2\vx^{~\!2}$ and  $\vx=\frac{\sqrt{2}}{\sqrt{3}\gamma_*^2}\,\vg_*$, where we have set $\gamma_*:=|\vg_*|$. 
Being on the boundary of the acceleration cone $\mathcal{A}$, cf.~\eqref{52}, this point has vanishing acceleration. 
 In addition we must have $\gamma_*^2=2\left(1+\frac{k}{\dot{a}^2}\right)^{-1}$ for the constraint \eqref{1stFried2} to be satisfied, which implies $\gamma^2_*>2$ for $k=-1$.  
This point  corresponds to {\it curvature scaling}. It merges with $P_0$, $P_2$,  for $\gamma_*\rightarrow\infty$, $\gamma_*\rightarrow\sqrt{2}$, respectively.

$\bullet$  The point $P_2=(\vx,z)$ with  $z^2=1-\vx^{~\!2}$ and $\vx=\frac{1}{\sqrt{6}}\,\vg_*$, with $\gamma_*^2< 6$.
Being on  the unit sphere ${\mathcal{S}}$, this point requires  $k=0$, and  corresponds to {\it scalar domination}. 
It merges with $P_{\mathcal{C}}$  for $\gamma_*\rightarrow\sqrt{6}$. The limit   $\gamma_*\rightarrow0$, in which case $P_2$  merges with the north pole of $\mathcal{S}$,  corresponds to constant potential and de Sitter universe.

\vfill\break

\renewcommand{\arraystretch}{2}
\begin{table}[H]
\begin{center}
\centering
\begin{tabular}{| c | c | c  | c |}
\hline
\cellcolor[gray]{0.9} Point & \cellcolor[gray]{0.9} $(\vx,z)$ &  \cellcolor[gray]{0.9} Interpretation &    \cellcolor[gray]{0.9} Conditions \\[4pt]
\hline
$P_{\mathcal{C}}$ & $(\vec{n},0)$  & kinetic domination & $k=0$, $V=0$ \\[7pt]
\hline
$P_0$ &  $(0,0,0)$  & curvature domination  &  $k=-1$, $V=0$ \\[7pt]
\hline
$P_1$ & $\frac{\sqrt{2}}{\sqrt{3}\gamma_*^2}(\vec{\gamma}_*,\sqrt{2}\gamma_*)$  &  curvature scaling  &  $k=-1$, $\gamma_*^2>2$ \\[7pt]
\hline
$P_2$ & $\frac{1}{\sqrt{6}}(\vec{\gamma}_*,\sqrt{6-\gamma_*^2})$  &  scalar domination & $k=0$, $\gamma_*^2<6$ \\[7pt]
\hline
\end{tabular}
\end{center}
\caption {Critical points of the dynamical system \eqref{system2} subject to the constraint \eqref{1stFried2}.~We list their physical interpretation,  their  coordinates in the $(\vx,z)$-space,  and necessary existence conditions.
The condition  $V=0$  for $P_{\mathcal{C}}$, $P_0$ need only hold asymptotically.~Additional existence conditions depend on the form of the potential.~We have restricted to $z\geq0$ and $k\leq0$; $\vg_*$ 
is the value of $\vg$ at the critical point of the system,  $\vec{n}$ is  an arbitrary unit vector,  and $\gamma_*:=|\vg_*|$. }
\label{tab:fixedpointsfields}
\end{table}
\renewcommand{\arraystretch}{1}

The physical interpretation of the different fixed points can  be seen from 
the analytic solutions of Section \ref{sec:2.3} below, and 
the corresponding energy balance between kinetic, potential, and curvature energy densities, 
$\Omega_{\text{kin}}$, $\Omega_{\text{pot}}$, and $\Omega_{c}$ respectively. More specifically,  we define, 
\eq{\label{427}
\rho_{\text{crit}}:=\frac{3H^2}{8\pi G}~;~~~
\Omega_{\text{kin}}:=
\frac{\frac12\dot{\vec{\varphi}}^{~\!2}}{\rho_{\text{crit}}}~;~~~
\Omega_{\text{pot}}:=\frac{V}{\rho_{\text{crit}}}
~;~~~
\Omega_{c}:=-\frac{k}{\dot{a}^2}
~,
}
so that,
\eq{\label{428}
\Omega_{\text{kin}}=\vec{x}^{~\!2}~;~~~\Omega_{\text{pot}}=z^2~,
}
and \eqref{1stFried2} reads, 
\eq{
\Omega_{\text{kin}}+\Omega_{\text{pot}}+\Omega_{c}=1
~.}
The resulting energy budgets are displayed in Table~\ref{table:energy}. 

\vfill\break

\renewcommand{\arraystretch}{2}
\begin{table}[H]
\begin{center}
\centering
\begin{tabular}{| c | c | c | c |}
\hline
\cellcolor[gray]{0.9} Point & \cellcolor[gray]{0.9} $\Omega_c$ &  \cellcolor[gray]{0.9} $\Omega_{\text{kin}}$ &  \cellcolor[gray]{0.9} $\Omega_{\text{pot}}$ \\[4pt]
\hline
$P_{\mathcal{C}}$ & 0 & 1 & 0 \\[7pt]
\hline
$P_0$ &  1  & 0 & 0 \\[7pt]
\hline
$P_1$ &  $1-\frac{2}{\gamma_*^2}$    & $ \frac{2}{3\gamma_*^2}$  & $ \frac{4}{3\gamma_*^2}$ \\[7pt]
\hline
$P_2$ &  0  & $\frac{\gamma_*^2}{6}$ & $1-\frac{\gamma_*^2}{6}$  \\[7pt]
\hline
\end{tabular}
\end{center}
\caption {Energy budget at the fixed points.~The critical points $P_{\mathcal{C}}$, $P_0$ are fully dominated by the kinetic, curvature energy respectively.~The energy 
budgets of $P_1$ and $P_2$ coincide for the limiting  value $\gamma_*=\sqrt{2}$, for 
which the curvature energy density of $P_1$ vanishes.}
\label{table:energy}
\end{table}
\renewcommand{\arraystretch}{1}

\begin{figure}[H]
\begin{center}
\includegraphics[width=.7\textwidth]{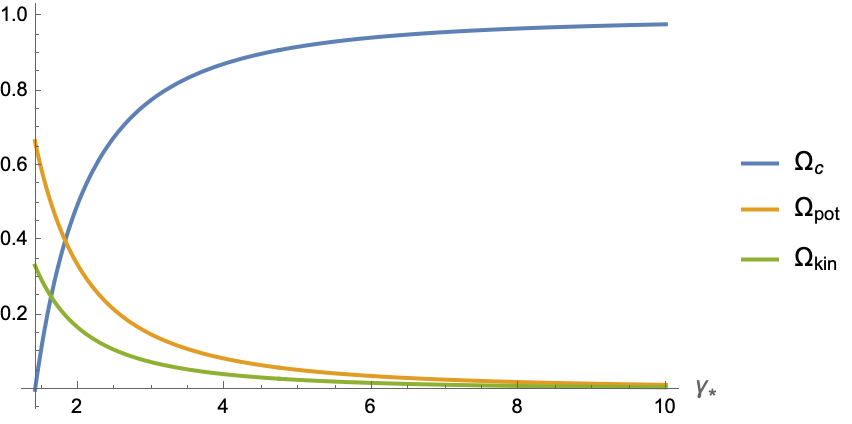}
\caption{Curvature, potential and kinetic energy densities  at the critical point  $P_1$, as functions of  the effective exponent $\gamma_*\geq\sqrt{2}$. The curvature energy density  vanishes  for $\gamma_*\rightarrow\sqrt{2}$, while it completely dominates for $\gamma_*\rightarrow\infty$. The ratio $\Omega_{\text{pot}}/\Omega_{\text{kin}}=2$ is  $\gamma_*$-independent.}\label{fig:engmix}
\end{center}
\end{figure}

The point $P_1$ will be of particular interest, as it is an  attractor for potentials typically coming from low-energy effective models  of string theory compactifications, i.e.~with steep exponentials ($\gamma_*^2>2$) in accordance
with the   swampland conjectures \cite{Obied:2018sgi,Hebecker:2018vxz,Andriot:2019wrs,Lust:2019zwm,Bedroya:2019snp,Andriot:2020lea,Rudelius:2021oaz,Rudelius:2021azq}.  
In Fig.~\ref{fig:engmix} we have plotted the dependence of the energy densities as a function of the effective exponent $\gamma_*$. The ratio of potential to kinetic energy at $P_1$
 is  $\gamma_*$-independent, 
$\Omega_{\text{pot}}/\Omega_{\text{kin}}=2$.




\vfill\break

\subsection{Analytic solutions}\label{sec:2.3}

To obtain the explicit solutions at the critical points, we 
start by integrating \eqref{variables2}, noting that $\frac{1}{H}\dot{\vv}=\vv^{~\!\prime}$, 
to obtain,  
\eq{\vec{\varphi}_*= \vec{\varphi}_{0}+\sqrt{6}\vx_*N~;~~~
\dot{\vec{\varphi}}_*= \sqrt{6}\vx_*H
~,}
for some constant vector $\vec{\varphi}_{0}$, where $\vx_*$ are the $\vx$-coordinates of the critical point.~For $P_\mathcal{C}$, $P_0$, for which $V=0$, plugging the above into the 
equations of motion  \eqref{2bgfieldeom}, \eqref{2bgF} immediately leads to the analytic solution. 

For the critical points $P_1$, $P_2$ we proceed by    evaluating \eqref{26} at the critical point  
to  obtain, 
\eq{\label{216}
V(\vv_*)=V_0e^{-\vg_*\cdot\vv_*}~,
}
 for some constant $V_0\geq0$. Inserting this expression for $V$ in the definition of $z$ in \eqref{variables2} and integrating, then leads to,  
 \eq{
 a_*(t)=a_0 ~\!t^{~\!\frac{\sqrt{2}}{\sqrt{3}\vg_*\cdot\vx_*}}~, 
 }
for some constant $a_0>0$, where  we have used the time-translation invariance to set $a_*(0)=0$. Plugging the above  into the equations of motion \eqref{2bgfieldeom}, \eqref{2bgF}, then  leads to certain additional algebraic conditions.  
The complete results of this analysis  are summarized in 
Table~\ref{tab:fixedpointssols}.~We see, in particular, that at each critical point, 
\eq{\label{218}
\vv_*(t)=\vv_0+c~\!\vx_*\ln t~,
}
with $c$  a proportionality factor  that depends on the critical point.

\renewcommand{\arraystretch}{2}
\begin{table}[H]
\begin{center}
\centering
\begin{tabular}{| c | c | c | c |}
\hline
\cellcolor[gray]{0.9} Point & \cellcolor[gray]{0.9} $a_*(t)$ &  \cellcolor[gray]{0.9} $\vv_*(t)$ &  \cellcolor[gray]{0.9} Conditions  \\[4pt]
\hline
$P_{\mathcal{C}} $ & $a_0\,t^\frac{1}{3}$  & $\vv_{0}+\vec{n}\sqrt{\frac23}\ln t$ & $k=0$, $V=0$   \\[2pt]
\hline
$P_0$ &  $t$  & $\vv_0$ & $k=-1$, $V=0$     \\[2pt]
\hline
$P_1$ & $\frac{~~\gamma_*}{\sqrt{\gamma_*^2-2}}\, t$  & $\vv_{0}+\frac{2}{\gamma_*^2}\vg_*~\!\ln t$  & \makecell{\\[-7pt]$k=-1$ \\[2pt] 
$\vec{\gamma}_*\cdot\vv_0=\ln  \frac{\gamma_*^2V_0}{4}$  \\ [-7pt]{}}   \\
\hline
$P_2$ & $a_0\, t^{\frac{2}{\gamma_*^2}}$  & $\vec{\varphi}_{0}+\frac{2}{\gamma_*^2}\vg_*~\!\ln t$ & \makecell{\\[-7pt]$k=0$ \\[2pt]
$\vec{\gamma}_*\cdot\vec{\varphi}_0=\ln\frac{\gamma_*^4V_0}{2(6-\gamma_*^2)}$\\ [-7pt]{}}   \\
\hline
\end{tabular}
\end{center}
\caption {Explicit solution for $a(t)$ and $\vv(t)$ at the critical  points of the dynamical system.~The condition  $V=0$ for $P_{\mathcal{C}}$, $P_0$ need only hold asymptotically.~Additional existence conditions depend on the form of the potential.~The quantities 
with a 0-subscript are free parameters, unless 
otherwise  indicated.~It is understood that $a_0>0$, and that $\vec{n}$ is an arbitrary unit vector. 
We have used the time-translation invariance to set $a_*(0)=0$.  }
\label{tab:fixedpointssols}
\end{table}
\renewcommand{\arraystretch}{1}

Crucially, the question of stability, and any  additional  existence conditions  of these critical  points,  depend on 
the dynamics of $\vg(\varphi)$, and thus on 
the exact form of $V$.~In particular, additional  existence conditions will generally be imposed by requiring that Eq.~\eqref{216} be satisfied. 
Indeed, this equation is not automatic, unless $V$ is a single exponential potential. 
We  examine these questions in the following sections, in the context of (multi-)exponential potentials.

\section{Single exponential potential}\label{sec:single}

One scalar theories with a single exponential potential have been studied extensively, see  \cite{Halliwell:1986ja} for the case of flat 3d space, and \cite{vandenHoogen:1999qq} for curved space with a  barotropic fluid.~Single exponential potentials with two fields and a barotropic fluid were discussed in  \cite{Guo:2003rs} for flat space, while   \cite{vandenHoogen:2000cf} also includes curvature and discusses the multi-field case.\footnote{The case of a single exponential 4d potential corresponds, from the standpoint of  the 
1d consistent truncation of \cite{Marconnet:2022fmx}, to the to the ``two-flux'' case therein.  Specifically, 
the system  \cite[Eq.~(61)]{Marconnet:2022fmx} reduces to  \eqref{system2} upon setting therein,  
\eq{
\spl{
\alpha_1\rightarrow-\gamma_1\sqrt{24}+24~;~~~ \beta_1\rightarrow&6~;~~~\gamma_1\rightarrow -\gamma_2\frac{1}{\sqrt{2}}\\
\alpha_2\rightarrow16~;~~~\beta_2\rightarrow &4~;~~~\gamma_2\rightarrow0
~,\nn}}
see also Section~\ref{sec:universal}.
 }

For a single exponential potential, 
\eq{\label{21exppot}
V(\varphi)=
  V_0 e^{-\vg\cdot\vv}~;~~~V_0\geq0
~,}
where  $\vg$   
and $V_0$ are constant,  
the dynamical  system \eqref{system2} is automatically  autonomous, so  no additional  conditions arise for the existence of the critical points discussed in Section \ref{sec:critical}. 
In this case one can in fact use a field redefinition  to decouple one of the two fields, 
and  essentially reduce the theory to  a single  scalar with an exponential potential \cite{Hartong:2006rt}.\footnote{\label{f:4}This follows from the fact that \eqref{2sc},\eqref{21exppot} are invariant under $\vv\rightarrow M\cdot\vv$, $\vg\rightarrow M\cdot\vg$, with $M\in O(2)$, which can be used to set e.g.~$\gamma_2=0$ in the new basis, thus decoupling the corresponding scalar $\varphi_2$.}   We will come back to this point in Section \ref{sec:contrunc}.

The stability properties of the critical points are determined by studying the eigenvalues and corresponding eigenvectors of  \eqref{system2} at each critical point.

$\bullet$  At $P_\mathcal{C}$, the dynamical system  has one positive eignenvalue (equal to 4) corresponding to a direction of approach along $\vn$. It has one vanishing eigenvalue 
corresponding to the direction of approach in the $\vx$-plane and perpendicular to $\vn$, which is a reflexion of the fact  that $\mathcal{C}$ is a circle of fixed points. 
The third eigenvalue is equal to $\sqrt{\frac32}(\sqrt{6}-\vg\cdot\vec{n})$,  
and corresponds to a direction of approach along $(0,0,1)$, i.e.~a trajectory on $\mathcal{S}$ approaching $P_\mathcal{C}$ along the $z$-direction. 
This eigenvalue is always positive  for $\gamma^2:=|\vg~\!|^2<6$. For  $\gamma^2>6$ the sign depends on the angle between $\vg$ and $\vn$.  We can understand this as follows: for $\gamma^2<6$ the fixed point 
$P_2$ exists and it is an attractor for trajectories on $\mathcal{S}$ (see below). In this case all trajectories on $\mathcal{S}$ start at a point of $\mathcal{C}$ end at $P_2$. For $\gamma^2>6$ the point $P_2$ does not exist, and trajectories on $\mathcal{S}$ starting 
at a point of $\mathcal{C}$ end at another point 
of $\mathcal{C}$. I.e.~in this case one segment of $\mathcal{C}$ consists of attractor points as concerns the approach along $(0,0,1)$, while the  points of the complementary segment are repulsive.

$\bullet$ At $P_0=(0,0,0)$, the dynamical system has one double negative eigenvalue (-2) corresponding  to a direction of approach along the $\vx$-plane. I.e.~$P_0$ is an attractor for all trajectories in the $\vx$-plane.~Indeed all these trajectories start at some point of $\mathcal{C}$ and end at $P_0$. The third eigenvalue is positive (+1) and corresponds to 
the unique trajectory (heteroclinic orbit) starting at $P_0$ and ending at $P_1$.

$\bullet$ At $P_1$ we have one negative eigenvalue (-2), corresponding to an approach parallel to the $\vx$-plane and perpendicular to $\vg$. 
The remaining two eigenvalues are 
$-1\pm\sqrt{8/\gamma^2-3}$, whose real part is always negative (recall that $\gamma^2>2$ for $P_1$ to exist). Moreover, if  $\gamma>\gamma_s$,  where,    
\eq{\label{scase}
 \gamma_s:=2\sqrt{\frac{2}{3}}~, 
 }
these eigenvalues are complex and $P_1$ is  a stable spiral, while for $\gamma\leq\gamma_s$ 
the eigenvalues are real and $P_1$ is  a stable node, see \cite{Andriot:2023wvg} for a recent discussion.

$\bullet$  At $P_2$ the system  has  one double  eigenvalue given by $(\gamma^2-6)/2$, which is negative  (recall that $\gamma^2<6$ for $P_2$ to exist), and corresponds to directions of approach tangent to the unit sphere  $\mathcal{S}$. 
The third eigenvalue is given by $\gamma^2-2$ and can have either sign. For  sufficiently flat  exponential 
potential ($\gamma^2<2$), i.e.~when $P_1$ does not exist, this eigenvalue is  negative.~In this case  $P_2$  is the only  fully stable critical point of the system.~Conversely, for  sufficiently steep exponential 
potential ($\gamma^2>2$),   it is $P_1$ which is the only  fully stable critical point.  

\begin{figure}[H]
\begin{center}
\includegraphics[width=.5\textwidth]{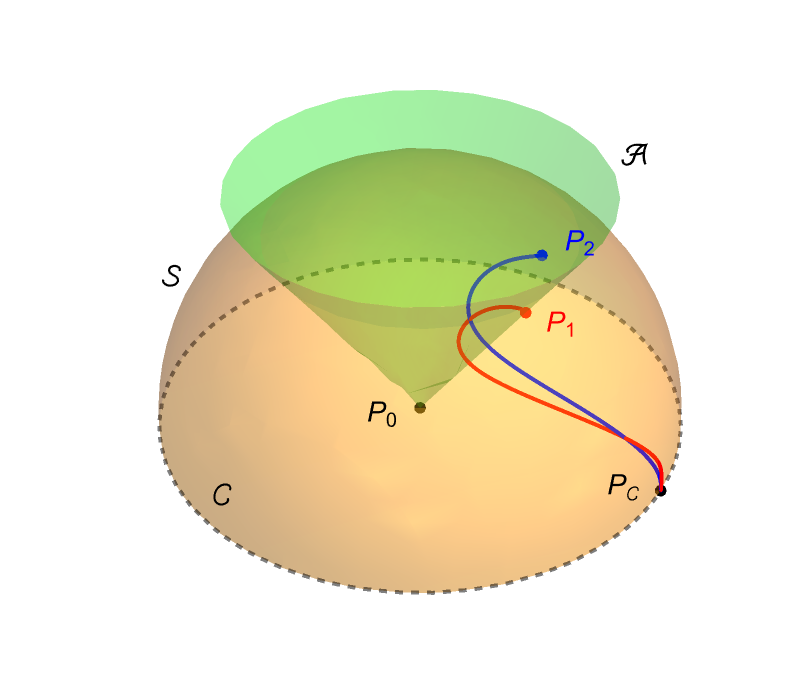}
\caption{\!Two trajectories in the interior of the unit sphere $\mathcal{S}$, asymptoting at past infinity the point $P_{\mathcal{C}}$ on the equator $\mathcal{C}$.~The blue trajectory asymptotes, at future infinity, the point $P_2$ with $\gamma^2=1.4$ on $\mathcal{S}$, and in the interior of the acceleration cone $\mathcal{A}$;~the red trajectory asymptotes the point $P_1$  with $\gamma^2=3.6$  on the boundary of $\mathcal{A}$, and in the interior of 
$\mathcal{S}$.}\label{fig_2a}
\end{center}
\end{figure}

The results of this analysis are summarized in Table~\ref{tab:fixedpointsstab}. One important conclusion is that  all open universes, which correspond to trajectories in the interior of the unit sphere $\mathcal{S}$, asymptote in the infinite past 
a fixed point  $P_\mathcal{C}$ dominated by kinetic energy. For sufficiently steep exponential potential ($\gamma^2>2$) they asymptote the point $P_1$  in the future, otherwise  (for $\gamma^2<2$) they asymptote $P_2$, cf.~Fig.~\ref{fig_2a}.

\vfill\break


\renewcommand{\arraystretch}{2}
\begin{table}[H]
\begin{center}
\centering
\begin{tabular}{| c | c | c | c |}
\hline
\cellcolor[gray]{0.9} Point & \cellcolor[gray]{0.9} $(\vx,z)$ &  \cellcolor[gray]{0.9} Eigenvalues &  \cellcolor[gray]{0.9} Conditions  \\[4pt]
\hline
$P_{\mathcal{C}} $ & $(\vec{n},0)$  & 4,\,0,\,$\sqrt{\frac32}(\sqrt{6}-\vg\cdot\vec{n})$ & — \\[7pt]
\hline
$P_0$ &  $(0,0,0)$  & $-2,\,1$  &  —  \\[7pt]
\hline
$P_1$ & $\frac{\sqrt{2}}{\sqrt{3}\gamma^2}(\vec{\gamma},\sqrt{2}\gamma)$  & $-2,\,-1\pm\frac{1}{\gamma}\sqrt{8-3\gamma^2}$  &  $\gamma^2>2$ \\[7pt]
\hline
$P_2$ & $\frac{1}{\sqrt{6}}(\vec{\gamma},\sqrt{6-\gamma^2})$  & $\frac12(\gamma^2-6)$, $\gamma^2-2$  & $\gamma^2<6$ \\[7pt]
\hline
\end{tabular}
\end{center}
\caption {Critical points of the single exponential potential:~coordinates, eigenvalues and existence conditions for $\vg$.~The point $P_1$ is fully stable whenever it exists (i.e.~for $\gamma>\sqrt{2}$), while $P_2$ is fully stable whenever the complementary 
condition holds ($\gamma<\sqrt{2}$).~The corresponding solutions at the critical points were given in Table~\ref{tab:fixedpointssols}.~All trajectories in the interior of $\mathcal{S}$ 
asymptote a point $P_\mathcal{C}$ at past infinity, and the fully stable point ($P_1$ or $P_2$) at future infinity.}
\label{tab:fixedpointsstab}
\end{table}
\renewcommand{\arraystretch}{1}

Due to the fact that $P_1$ lies on the boundary of 
the acceleration cone $\mathcal{A}$,  cosmologies  that asymptote   $P_1$  in the future always feature a period of accelerated expansion (which can be transient, eternal or semi-eternal). Moreover,  as emphasized in \cite{Marconnet:2022fmx, Andriot:2023wvg}, the  number of e-folds 
 can be parametrically controlled, by adjusting the distance of the trajectory to  the origin $P_0$ of the phase space, cf.~\cite[Figs.~\!7,8]{Andriot:2023wvg}.

\subsection{One-field consistent truncation}\label{sec:contrunc}

As already mentioned, for a single exponential potential one  of the two scalar fields can be decoupled by a field redefinition. 
This can also be seen directly from the dynamical system: 
with $\vg$ being constant in the single exponential case, 
an additional  consequence of  \eqref{system2} is that the plane, 
\eq{\mathcal{P}:=\left\{(\vx, z)\in\mathbb{R}^3~|~x_1\gamma_2-x_2\gamma_1=0\right\}
~,}
is an invariant surface. Indeed from \eqref{system2}  we have, 
\eq{\label{explain}
x_1'\gamma_2-x_2' \gamma_1=
-\left[z^2+2 (1 - x_1^2 - x_2^2) \right] (x_1\gamma_2-x_2 \gamma_1)
~.}
It follows from the above and the definition \eqref{variables2} of $\vx$, that restricting to trajectories on $\mathcal{P}$ is equivalent  to setting  
$\varphi_1\gamma_2-\varphi_2\gamma_1$  equal to an arbitrary  constant, 
thereby consistently truncating out one of the two scalars.~Assuming $\gamma_1\neq 0$ without loss of generality, this consistent truncation is thus obtained by setting,  
\eq{
\varphi_2=\frac{\gamma_2}{\gamma_1}\varphi_1+\text{const.}~;~~~\varphi:=\frac{\gamma}{\gamma_1}\varphi_1
~,}
where $\gamma:=|\vg|$, 
upon which the theory \eqref{2sc} truncates to, 
\eq{\label{2sctrunc}
S_{4\d}=\int\d^4 x\sqrt{g}\left(
\frac{1}{16\pi G}R-\frac12 g^{\mu\nu}   \partial_\mu \varphi\partial_\nu \varphi
  -\tilde{V}_0 e^{-\gamma\varphi}
\right)~,}
for some constant $\tilde{V}_0\geq0$, in general different from $V_0$ of \eqref{21exppot}.~In particular we see that the one-field truncation can only increase the effective exponent of the potential, 
since $|\vg|\geq|\gamma_1|,|\gamma_2|$.

The resulting model has a two-dimensional  phase space, parameterized by  $z$, as given  in  \eqref{variables2}, and $x$ defined by, 
\eq{\label{variables1}
x := \frac{\dot{\varphi}}{H \sqrt{6}} ~,
}
so that \eqref{system2}, \eqref{1stFried2} remain valid, provided we simply replace $\vx$ by $x$.

\subsection{Heteroclinic orbit: unique analytic solution}   \label{sec:heteroc}

Pure  higher-dimensional gravity compactified on Einstein manifolds  was considered in \cite{Andersson:2006du}.~In particular, the cosmological solutions of the 10d model compactified on a 6d hyperbolic space were shown to exhibit  late-time acceleration. In \cite{Marconnet:2022fmx} we pointed out that these models admit cosmological solutions with transient acceleration and parametric control of e-folds.

We will now show that 
the cosmological solution corresponding to the unique trajectory (called a {\it heteroclinic orbit} in dynamical system terminology) asymptoting the critical point $P_0$ in the  past and  the critical point $P_1$ in the future can be given analytically in the special case $\gamma=\gamma_s$,  the threshold value  above (below) which $P_1$ is a stable spiral (node), cf.~Eq.~\eqref{scase}.~As was  noted  in  \cite{Marconnet:2022fmx}, the critical point $P_0$ corresponds to a smooth 4d geometry (a Milne universe)  that  is reached at finite proper time in the past.~The orbit can thus be geodesically completed beyond $P_0$ to $t<0$, using the time-reversal symmetry of Eqs.~\!\eqref{2bgfieldeom},~\!\eqref{2bgF}:~it corresponds to an eternally accelerating solution without Big Bang singularity.\footnote{\samepage As 
pointed out in \cite{Marconnet:2022fmx}, in the vicinity of $t=0$,  the solution describes a de Sitter space in hyperbolic slicing, 
\eq{\d s^2=-\d t^2+\sinh^2(t) \left(\frac{\d r^2}{1 + r^2} + r^2 \d \Omega^2 \right)
~.\nn}
 Indeed, as it can be seen directly from Footnote~\ref{330},   
 $a(t)=\sinh(t)$, 
 up to and including $\mathcal{O}(t^3)$ terms. 
}

 We first note that for  general $\gamma$ we can obtain the solution of the 
  unique trajectory asymptotically in the neighborhood of the origin of phase space $x=z=0$. 
Indeed,  from \eqref{system2}, we obtain a perturbative expansion for $x$, $z$ in terms of $N$, 
 \eq{\spl{\label{xyana}
 \frac{\sqrt{6}}{\gamma}x(N)&=\frac34 {a}^2 -\frac{1}{2^4} (10+3\gamma^2) {a}^4
  +\frac{5}{2^{10}} (112+78\gamma^2+9\gamma^4) {a}^6+ 
\dots
\\
 z(N)&=  {a} -\frac{1}{2^4} (8+3\gamma^2) {a}^3
  +\frac{1}{2^9} (192+160\gamma^2+21\gamma^4) {a}^5+ 
\dots
 ~,}} 
 where  $a= e^N$,  
  so that $(x,z)$ tends to the origin of phase space ($P_0$) asymptotically in the past,  $N\rightarrow-\infty$, $a\rightarrow0$, in a direction of approach along the $z$-axis.\footnote{
  Explicitly, we substitute  $x=\sum_{i=1}^\infty\varepsilon^i\delta x^{(i)}$, $z=\sum_{i=1}^\infty\varepsilon^i\delta z^{(i)}$ into  \eqref{system2}, and we solve for $\delta x^{(i)}, \delta z^{(i)}$ order-by order in $\varepsilon$. 
The approach along the $z$-axis is imposed by setting $\delta x^{(0)}=0$.  The auxiliary parameter   $\varepsilon$ may be set to one at the end. 
  } 
  These expansions can be straightforwardly obtained to any desired order in  $a$.

As it turns out, in the special case $\gamma=\gamma_s$,   from the series expansions  
\eqref{xyana} we can guess the closed expressions, 
 \eq{\label{xyan}
 x=\frac12\left(1-\frac{1}{
 \sqrt{1+2a^{2}}
 }
 \right)
 ~;~~~
 z=
\frac{a}{
 \sqrt{1+2a^{2}}
 }
~,} 
which, as we verify a posteriori, correspond to bona fide solutions of the equations of motion (see below). 
Moreover, asymptotically in the   future,~${N}, a\rightarrow\infty$, the solution tends to, 
 \eq{
(x,z)= ( \frac{1}{2} , \frac{1}{\sqrt{2}} )
 ~,}
 which are indeed the coordinates of $P_1$ for the special case \eqref{scase}, cf.~Table~\ref{tab:fixedpointsstab}.~I.e.~although we obtained the solution by initially expanding around  
$P_0$, the analytic solution \eqref{xyan} remains valid for the entire length of the orbit from $P_0$ to $P_1$. 
 
 In order to obtain the analytic solution in terms of   the original variables of the cosmological model, $\varphi(t)$, $a(t)$,  
 we proceed as follows. 
Integrating \eqref{variables2}, taking \eqref{xyan} into account, we  find,
%
%
%
%
\eq{\spl{\label{phian}
\varphi&=\varphi_{-\infty} +
 \sqrt{\frac32} ~\text{ln}\left(    \frac{ 1+\sqrt{1+2a^{2}}  }{ 2 }   \right)
~,}}
where  the constant $\varphi_{-\infty}$ is given by, 
\eq{\label{phi0}
e^{2\sqrt{\frac23}\varphi_{-\infty}}= \frac13 V_0 
~,}
so that 
for $N\rightarrow-\infty$, $a\rightarrow 0$, we have $\varphi\rightarrow\varphi_{-\infty}$.  
Moreover, integrating the constraint \eqref{1stFried2} we obtain,
\eq{\spl{\label{tan}
t&= \frac12~\! a+\frac{1}{2\sqrt{2}}~\!\text{arcsinh}(\sqrt{2}~\!a)
~,}}
where we have imposed, without loss of generality, that $t\rightarrow0$ as $a\rightarrow 0$.  
Eq.~\eqref{tan}  implicitly 
 defines $a(t)$, and thus also $\varphi(t)$, via \eqref{phian}.\footnote{\label{330} 
 The dilaton and the scale factor   can be given  explicitly as  functions of cosmological time,  for any  $\gamma$,  and to any desired order in a perturbative expansion 
 around $t=0$, 
\eq{\spl{
a&=t+\frac16 ~\!t^3+\frac{1}{15\cdot2^{6}}(8-27\gamma^2)~\!t^5+\mathcal{O}(t^7)\\
 \frac{1}{\gamma}(\varphi- \varphi_{-\infty})&= \frac38  ~\!t^2 - 
 \frac{1}{2^6}  (2 + 3 \gamma^2) ~\!t^4 + \frac{1}{15\cdot2^{11}} (112 + 
    342 \gamma^2 + 225\gamma^4)~\! t^6+\mathcal{O}(t^8)~.\nn
}}
 However we have been unable to find a closed expression for the right-hand sides of the equations  above, even for the spacial case $\gamma=\gamma_s$.}





\begin{figure}[H]
\begin{center}
\includegraphics[width=.5\textwidth]{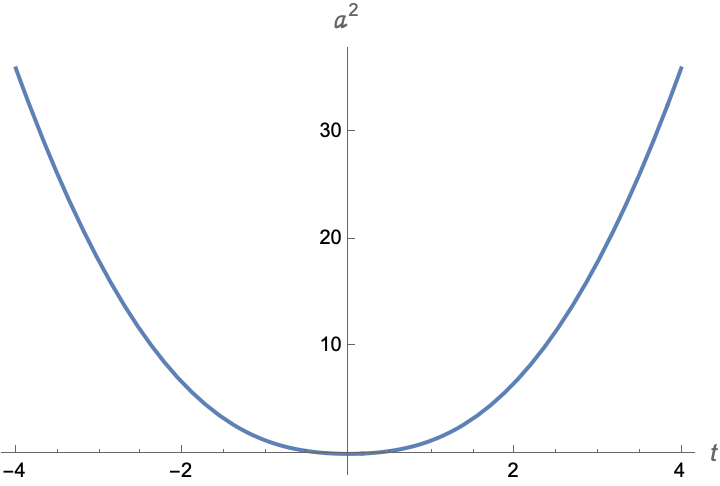}
\caption{Plot of the scale factor of the heteroclinic orbit as a function of cosmological time. The cosmology is accelerating for all $t$; it is expanding (contracting) for $t>0$ ($t<0$).~The 4d geometry remains smooth at the apparent singularity $a= 0$, in the vicinity of which spacetime is 
quasi de Sitter.}\label{heteroc}
\end{center}
\end{figure}

We can  directly verify that the analytic solution \eqref{phian}-\eqref{tan}   satisfies the  equations of motion  \eqref{2bgfieldeom},~\eqref{2bgF}.~To that end it suffices to  use,  
\eq{\label{322}
H=\frac{\d N}{\d t}=
 \frac{2 \sqrt{1+2a^2}  }{ a \left(1+\sqrt{1+2a^{2}}  \right)} 
~,}
which follows from \eqref{tan}, in order to convert derivatives with respect to $t$ to derivatives with respect to $N$. 
We can also verify that in the limit $t\rightarrow0$ ($t\rightarrow\infty$), the solution  \eqref{phian}-\eqref{tan}  asymptotes  the analytic solution at the critical point 
$P_0$ ($P_1$) given in Table~\ref{tab:fixedpointssols}.

The energy densities \eqref{427} along the heteroclinic orbit  can also be given analytically,
\eq{
\Omega_{\text{pot}}=\frac{ a^{2}  }{ 1+2a^{2}    }~;~~~
\Omega_{\text{kin}}=\frac{\left( -1+\sqrt{1+2a^{2}}   \right)^2}{4\left(  1+2a^{2}    \right)}~;~~~
\Omega_{c}=\frac{\left( 1+\sqrt{1+2a^{2}}   \right)^2}{4\left(  1+2a^{2}    \right)}
~.}
In the asymptotic future ($P_1$) we have $\Omega_{\text{pot}}\rightarrow\frac12$, and $\Omega_{\text{kin}},\Omega_{c}\rightarrow\frac14$. In the asymptotic past ($P_0$) we 
have $\Omega_{c}\rightarrow 1$, and $\Omega_{\text{kin}},\Omega_{\text{pot}}\rightarrow0$, in accordance  with Table~\ref{table:energy} for  $\gamma_*=\gamma_s$.

\vfill\break




\begin{figure}[H]
\begin{center}
\includegraphics[width=.7\textwidth]{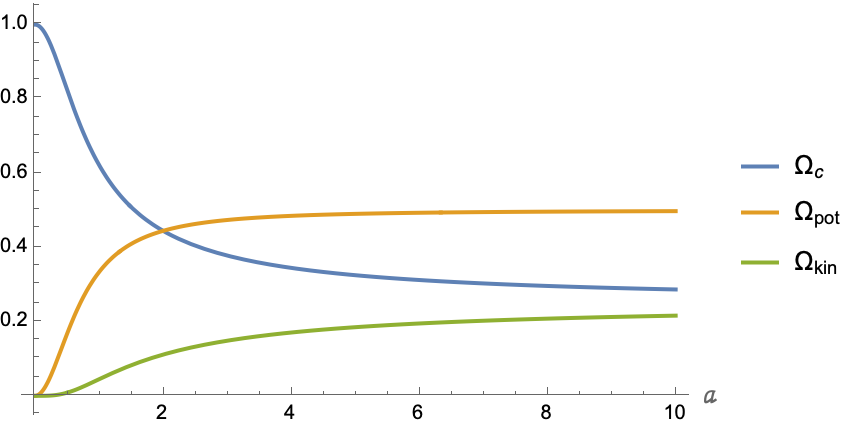}
\caption{The evolution of the energy densities along the heteroclinic orbit from $P_0$, in the past  ($a\rightarrow0$), to $P_1$ in the asymptotic future ($a\rightarrow\infty$).~The energy densities at the endpoints of the orbit asymptote those for   $P_0$, $P_1$  given in Table~\ref{table:energy} for  $\gamma_*=\gamma_s$.}\label{fig2}
\end{center}
\end{figure}

\section{Two-exponential potential}\label{sec:2exp}

Let us now consider a two-exponential potential with ``exponents'' $\va, \vb$, 
\eq{\label{2exppot2}
V(\varphi)=
  A e^{-\va\cdot\vv}+
    B e^{-\vb\cdot\vv}
~,}
where  
 $A,B$,  are real constants.~We will assume $\va\neq\vb$, otherwise we are  back to the single exponential potential case discussed in Section~\ref{sec:single}.~For the reasons already mentioned in Footnote~\ref{f1},  we are interested in non-negative potentials, so we may assume $A\geq0$ without loss of generality, while allowing 
 $B$ to have either sign.
 
 Two-exponential potentials with one scalar field were discussed in  \cite{Barreiro:1999zs,Li:2005ay,Jarv:2004uk}. 
 Multi-exponential potentials with a different scalar field in each exponential were considered in  \cite{Liddle:1998jc,Coley:1999mj}. This is the setup of what is usually referred to as {\it assisted inflation}, and corresponds, in our notation,  to 
having $\va\cdot\vb=0$.\footnote{To see this, use the invariance discussed in Footnote~\ref{f:4} so that in the new basis e.g.~$\va\propto(1,0)$ and $\vb\propto (0,1)$.} 
The case of multi-exponential potentials with multiple fields in each exponential, called   {\it generalized assisted inflation} in the literature,  was discussed in \cite{Copeland:1999cs,Collinucci:2004iw} for flat 3d space, while  partial results concerning 
non-vanishing 3d  curvature were given in \cite{Hartong:2006rt}.

The system  \eqref{system2},\eqref{1stFried2}  now has to be supplemented with the 
dynamical equations for $\vg(\varphi)$,
\eq{\label{geom}
\gamma'_i
=\sum_{j=1}^2\sqrt{6}x_j\left(
\gamma_i\gamma_j-\gamma_{ij}
\right)
~,}
where we  defined  \cite{Copeland:1997et,Steinhardt:1999nw},
\eq{
\gamma_{ij}(\varphi):=\frac{\partial_{\varphi_i}\partial_{\varphi_j}V}{V}
~.}
For the potential \eqref{2exppot2}, 
  the parenthesis on  right hand side of \eqref{geom}  is given  in terms of the $\gamma_i$'s,
\eq{\label{44}
 \gamma_i\gamma_j-\gamma_{ij}=( \alpha_i- \beta_i) ( \alpha_j- \beta_j) 
\frac{
( \alpha_1  \gamma_2- \alpha_2  \gamma_1) ( \beta_1
    \gamma_2- \beta_2  \gamma_1)
    } {
 (   \alpha_1\beta_2-\alpha_2\beta_1)^2}
~,}
hence the system  \eqref{system2}, \eqref{geom} is autonomous. 
We may then rewrite \eqref{geom},\eqref{44} as, 
\eq{ \label{geom2} 
\vg^{~\!\prime}=\frac{\sqrt{6}
( \va\wedge  \vg)\cdot ( \vb\wedge
    \vg)
    } {
 |\va\wedge \vb|^2} \,
\left[(\va-\vb)\cdot\vec{x}\right](\va-\vb)
~,}
where the vector products among the two-component vectors $\va,\vb,\vg$ in  \eqref{geom2} should be thought of as embedded in an auxiliary three-dimensional space.~This is simply a calculational trick that allows us 
to rewrite \eqref{geom} in  vector notation.

The dynamical system \eqref{system2}, \eqref{geom2}, subject to the constraint \eqref{1stFried2}, is thus five-dimensional: cosmologies now correspond to 
trajectories in $(\vx,z,\vg)$-space.~These  phase space variables are different from  the so called ``EN variables'' \cite{Copeland:1997et} typically used in multi-exponential systems \cite{we, Bahamonde:2017ize}.~Moreover, the projection of the motion onto the $\vg$-plane admits a simple, intuitive description.~Indeed, 
as follows from its definition and the form of the potential, $\vg$ is constrained to move on the line connecting $\va$ and $\vb$.\footnote{This is also 
consistent with \eqref{geom2}, as can be seen from the fact that the right-hand side of that equation is proportional to $(\va-\vb)$.}

\begin{figure}[H]
\begin{subfigure}{.5\textwidth}
  \centering
  \includegraphics[width=0.5\linewidth]{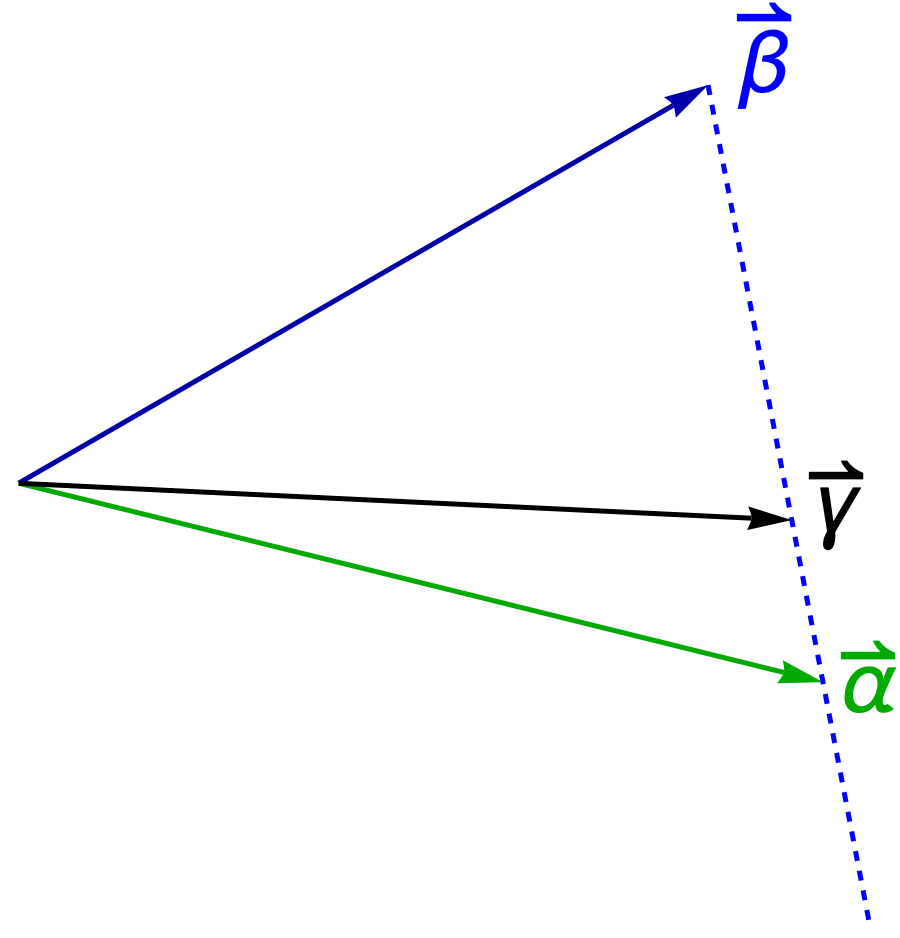}
  \caption{The case $A,B>0$: $\vg$ is constrained to lie\\ on the segment  between the two vectors $\va,\vb$.\\{}}
  \label{fig8a}
\end{subfigure}%
\begin{subfigure}{.5\textwidth}
  \centering
  \includegraphics[width=0.5\linewidth]{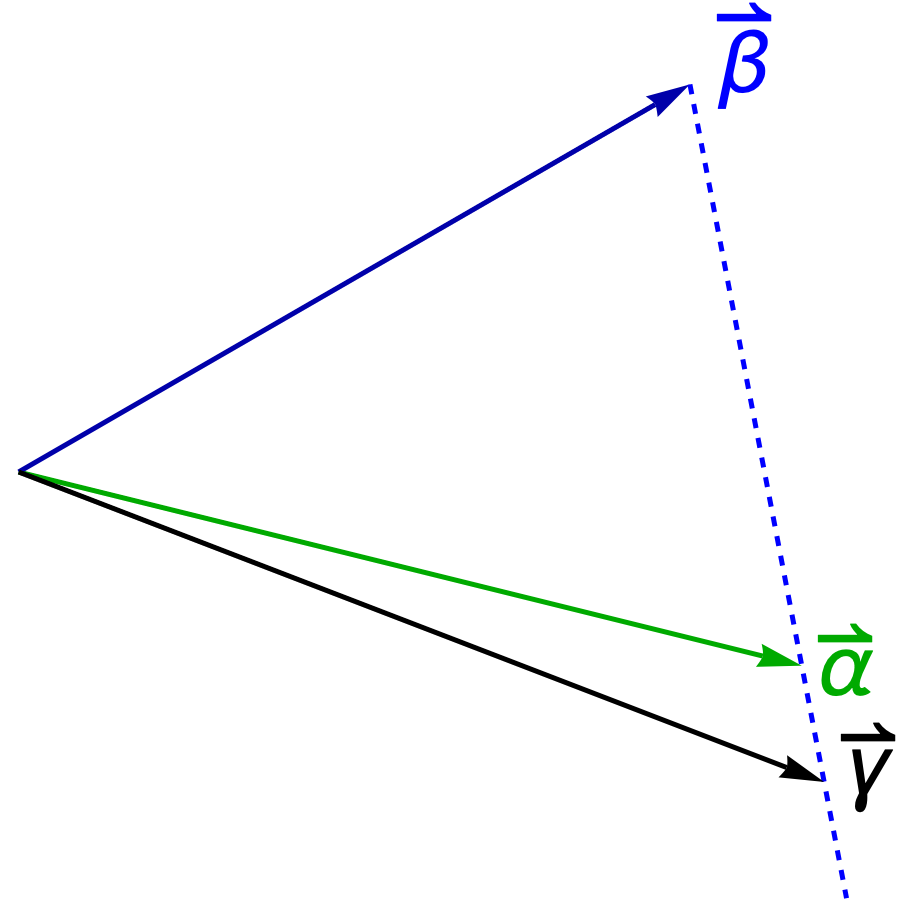}
  \caption{The case $A>0,B<0$: $\vg$ is constrained to lie on the exterior of the segment between the two vectors $\va,\vb$, and on the side of $\va$.}
  \label{fig8b}
\end{subfigure}
\caption{Projection of motion onto the $\vg$-plane.~Restricting, without loss of generality, to the case $A>0$, $\vg$ is constrained to lie on the semi-infinite line joining the two vectors $\va,\vb$, which starts at $\vb$ and extends to the side of $\va$ (dotted blue line).
}
\label{fig8}
\end{figure}

To see this, let us first consider the case $A,B>0$. 
Without loss of generality we can absorb these two constants by shifting $\vv$ by a constant vector, so that,  
\eq{\label{470}
\vg=\mu\va+(1-\mu)\vb~;~~~\mu:=\frac{e^{-(\va-\vb)\cdot\vv}}{e^{-(\va-\vb)\cdot\vv}+1}\in(0,1)
~,}
therefore $\vg$ is constrained to move on the segment of the line between the two vectors $\va,\vb$, cf.~the black arrow in Fig.~\ref{fig8a}.

Let us now consider the case where $A,B$ are not both positive: $A>0$, $B<0$. Proceeding as in the previous case, we may absorb these constants in the definition of $\vv$,  to obtain, 
\eq{  
V=e^{-\vb\cdot\vv}\left(e^{-(\va-\vb)\cdot\vv}-1\right)~.
} 
The requirement that $V>0$ thus constrains $e^{-(\va-\vb)\cdot\vv}\in(1,\infty)$, so that 
\eq{\label{471}
\vg=\mu\va+(1-\mu)\vb~;~~~\mu:=\frac{e^{-(\va-\vb)\cdot\vv}}{e^{-(\va-\vb)\cdot\vv}-1}\in(1,\infty)
~.}
We see that $\vg$ is constrained to move on the exterior of the segment between the two vectors $\va,\vb$, and on the side of $\va$, cf.~the black arrow in Fig.~\ref{fig8b}. 
%
%

The division of the line into three segments with boundaries given by $\va,\vb$, 
is consistent with the fact that  $\vg=\va,\vb$ are critical points of the system \eqref{geom2}.~This ensures that $\vg$ can only reach the boundary of a segment asymptotically.~In particular, 
$\vg$ cannot cross over to a segment that would violate the $V\geq0$ condition.

\subsection{Critical Points}\label{sec:41}

The critical points of the five-dimensional system \eqref{system2}, \eqref{geom2}, subject to the constraint \eqref{1stFried2}, must now, in addition to the conditions discussed in Section~\ref{sec:critical}, 
also satisfy $\vg^{~\!\prime}=0$. In view of the right-hand side of \eqref{geom2}, one way this condition is solved is if $\vg$ is parallel to $\va$ or $\vb$. 
Moreover, since $\vg$ is constrained to move on the line connecting $\va$ and $\vb$,  this is equivalent to, 
\eq{
\vg_*=\va~~\text{or} ~~\vg_*=\vb~,}  
which  means that the potential is dominated by one of the two exponentials.~Therefore  this case  reduces to the one 
already discussed in Section~\ref{sec:critical}, and  no additional conditions  arise, other than $A>0$ (for the case $\vg_*=\va$) or $B>0$ (for the case $\vg_*=\vb$), ensuring that the potential remains positive at the critical point.

The remaining possibility for the  right-hand side of \eqref{geom2} to vanish, is that $\vx$ is perpendicular to $(\va-\vb)$ at the critical point.~This is equivalent to the condition  
that both exponentials scale in the same way at the critical point, as can be seen from \eqref{218},\eqref{2exppot2}.~In the following will therefore impose, 
\eq{\label{cgi}
\vx_*\cdot(\va-\vb)=0~.}
Let us now examine the implications of \eqref{cgi} for each of the critical points of Table~\ref{tab:fixedpointsfields}:

$\bullet$ At $P_\mathcal{C}$ we have $\vx_*=\vn$, and  the condition \eqref{cgi} is not satisfied, in general, except for 
 the two antipodal points of the equator for which $\vn$ is perpendicular to $(\va-\vb)$ — in which case  $\vg_*$ is unconstrained.~This can also be 
seen directly from \eqref{216}, which then reduces to, 
\eq{\label{410}\vn\cdot\va=\vn\cdot\vb=\vn\cdot\vg_*~,}
which is simply  the statement that  $\vg_*$ can lie anywhere on the line connecting  $\va$ and $\vb$.

$\bullet$ At $P_0$ the conditions \eqref{cgi}, \eqref{216}  are automatically statified,  therefore $\vg_*$ is unconstrained.

\begin{figure}[H]
\begin{subfigure}{.5\textwidth}
  \centering
  \includegraphics[width=0.5\linewidth]{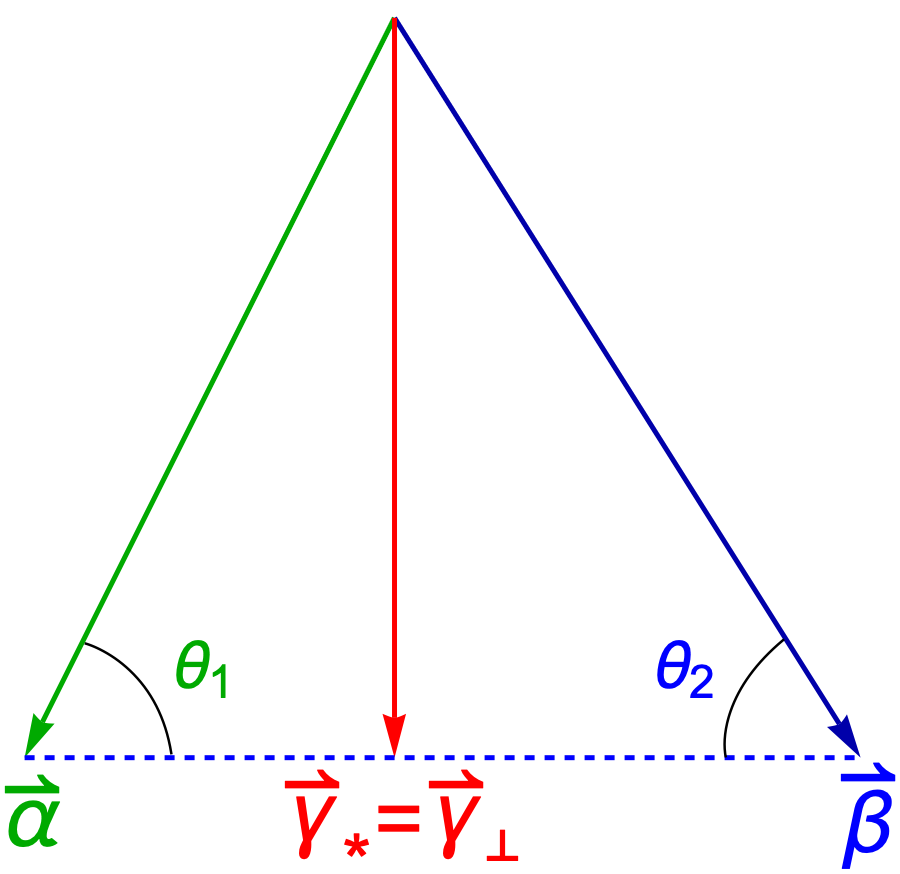}
  \caption{The case $\theta_1,\theta_2<\frac{\pi}{2}$: $\vg_*=\vg_\perp$ is stable;\\ it is in the  allowed range of $\vg$ iff $A,B>0$.}
  \label{fig4}
\end{subfigure}%
\begin{subfigure}{.5\textwidth}
  \centering
  \includegraphics[width=0.5\linewidth]{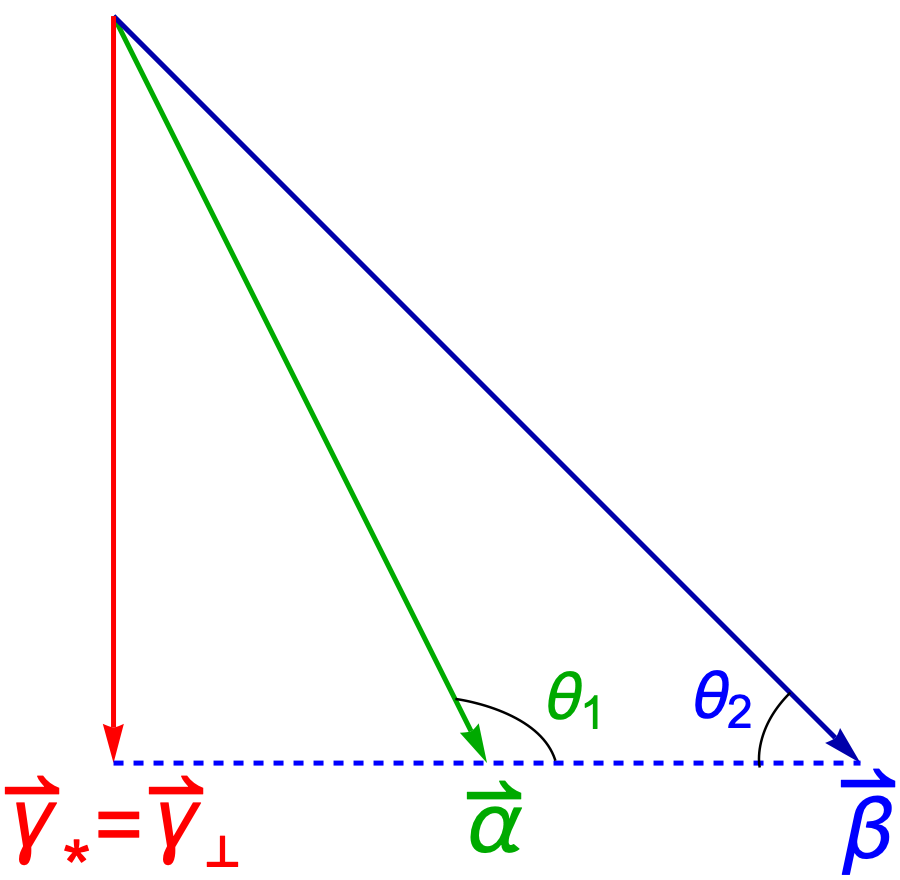}
  \caption{The case $\theta_1>\frac{\pi}{2},\theta_2<\frac{\pi}{2}$:  $\vg_*=\vg_\perp$ is unstable;\\ it is in the allowed range of $\vg$ iff $A>0,B<0$.}
  \label{fig3b}
\end{subfigure}
\caption{$\vg_*=\vg_\perp$ at the critical points $P_1$, $P_2$,  as given in \eqref{465}: it corresponds to the height of the triangle formed by $\va,\vb$; we are assuming $A>0$, without loss of generality. 
The conditions \eqref{dn}, \eqref{dn2} are satisfied iff $\vg_\perp$ is within the allowed range of $\vg$,   cf.~Fig.~\ref{fig8}. 
}
\label{fig34}
\end{figure}

$\bullet$ At $P_1$ the conditions \eqref{cgi}, \eqref{216}  reduce to, 
 \eq{\label{411}\vec{\alpha}\cdot\vec{\gamma}_*= \vec{\beta}\cdot\vec{\gamma}_*={\gamma}_*^2~,}
which  
 can be solved to determine $\vg_*$ in terms of $\va$, $\vb$,\footnote{It can be checked that in the $\va\rightarrow\vb$  limit,  \eqref{465} remains well-defined and reduces to $\vg_*=\va=\vb$.}
 \eq{\label{465}
 \vg_*=\vec{\gamma}_\perp:=\frac{(\va-\vb)\wedge(\vb\wedge\va)}{|\va-\vb|^2}~;~~~
  \gamma_*=|\vec{\gamma}_\perp|=\frac{|\va\wedge\vb|}{|\va-\vb|}
~.}
This is the statement that $\vg_*$ is the height, $\vg_\perp$, of the triangle  formed by $\va$, $\vb$, cf.~Figure~\ref{fig34}. 
%
%

Finally, 
inserting the solution for $a_*(t)$, $\vv_*(t)$ at $P_1$ from Table~\ref{tab:fixedpointssols}, with $\vg_*$ as given in  \eqref{465},  into the equations of motion \eqref{2bgfieldeom}, \eqref{2bgF},  we 
obtain two additional conditions,%
 \eq{\label{dn}
\vec{\alpha}\cdot\vec{\varphi}_0=\ln\left( \frac{A\,|\va\wedge\vb|^2}{4\vec{\beta}\cdot(\vec{\beta}-\va)}\right)~;~~~
\vec{\beta}\cdot\vec{\varphi}_0=\ln\left(  \frac{B\,|\va\wedge\vb|^2}{4\va\cdot(\va-\vb)}\right)
~.}
 It can readily  be checked, taking  \eqref{216},\eqref{465} into account,  that  
these imply the condition 
$\vec{\gamma}_*\cdot\vv_0=\ln  \frac{\gamma_*^2V_0}{4}$  from Table~\ref{tab:fixedpointssols}. 

Eqs.~\eqref{dn}  admit a solution, provided the arguments of the logarithms in the equations above are positive. 
 Specifically, $A,B$ must have the same sign as $\vec{\beta}\cdot(\vec{\beta}-\va)$, $\va\cdot(\va-\vb)$, respectively. 
The latter scalar products having positive (negative) sign  is equivalent to the angle $\theta_2$, respectively $\theta_1$, of the base of the triangle formed 
by $\va,\vb$ being smaller (greater) than $\frac{\pi}{2}$, cf.~Fig.~\ref{fig34}. 
 
For $A,B>0$, the conditions \eqref{dn} are thus equivalent to $\theta_1,\theta_2<\frac{\pi}{2}$. Equivalently, the height of the triangle must lie inside  the segment joining the two vectors $\va,\vb$, cf.~Fig.~\ref{fig4}.~On the other hand, this is precisely the allowed range of $\vg$ for $A,B>0$, cf.~Fig.~\ref{fig8}. 

For $A>0, B<0$, the angle $\theta_1$   must be greater than $\frac{\pi}{2}$ for \eqref{dn} to be satisfied. Equivalently,  the height of the triangle must lie on the line joining the two vectors 
$\va,\vb$, at a point exterior to the  segment
between the two vectors, and on the side of the vector $\va$, cf.~Fig.~\ref{fig3b}.~Again this precisely coincides with the allowed range of $\vg$, cf.~Fig.~\ref{fig8}. 

We conclude that, in every case, the conditions \eqref{dn} admit a solution if and only if $\vg_\perp$ lies within the allowed range of $\vg$.

$\bullet$ At $P_2$ the  same reasoning as in the case  of $P_1$   leads to the exact same conditions for $\vg_*$,  given in \eqref{465}.
Moreover, 
inserting the solution for $a_*(t)$, $\vv_*(t)$ at $P_2$ from Table~\ref{tab:fixedpointssols}   into the equations of motion \eqref{2bgfieldeom}, \eqref{2bgF},  we 
obtain two additional conditions,%
 \eq{\label{dn2}
\vec{\alpha}\cdot\vec{\varphi}_0=\ln\left(\frac{\gamma_*^4A}{2(6-\gamma_*^2)}~\!\frac{|\va-\vb|^2}{\vec{\beta}\cdot(\vec{\beta}-\va)}\right)~;~~~
\vec{\beta}\cdot\vec{\varphi}_0=\ln\left(\frac{\gamma_*^4B}{2(6-\gamma_*^2)}~\!\frac{|\va-\vb|^2}{\va\cdot(\va-\vb)}\right)
~.}
It can  be checked   that  these  imply the condition 
$\vec{\gamma}_*\cdot\vec{\varphi}_0=\ln\frac{\gamma_*^4V_0}{2(6-\gamma_*^2)}$   from Table~\ref{tab:fixedpointssols}.

Eqs.~\eqref{dn2}  admit a solution,  provided the arguments of the logarithms in the equations above are positive. 
 This leads to exactly the same conditions as discussed below \eqref{dn}, 
since $\gamma_*^2<6$, by the requirement of existence of $P_2$. 
We thus conclude that the conditions \eqref{dn2} admit a solution if and only if $\vg_\perp$ lies within the allowed range of $\vg$.

The results of this analysis are summarized in Table~\ref{tab:4}.

 \vfill\break

\renewcommand{\arraystretch}{2}
\begin{table}[H]
\begin{center}
\centering
\begin{tabular}{| c | c | c | c |}
\hline
\cellcolor[gray]{0.9} Point & \cellcolor[gray]{0.9} $a_*(t)$ &  \cellcolor[gray]{0.9} $\vv_*(t)$ &  \cellcolor[gray]{0.9} Conditions  \\[4pt]
\hline
$P_{\mathcal{C}}$ & $a_0\,t^\frac{1}{3}$  & $\vv_{0}+\vec{n}\sqrt{\frac23}\ln t$ & \makecell{\\[-7pt]$k=0$, $V=0$\\ $\vn\cdot(\va-\vb)=0$ \\[-7pt]{}}  \\
\hline
$P_0$ &  $t$  & $\vv_0$ & $k=-1$, $V=0$     \\[2pt]
\hline
$P_1$ & $\frac{~~\gamma_*}{\sqrt{\gamma_*^2-2}}\, t$  & $\vv_{0}+\frac{2}{\gamma_*^2}\vg_*~\!\ln t$  & \makecell{\\[-7pt]$k=-1$, $\vg_*=\vg_\perp$ \\[2pt]
$\vec{\alpha}\cdot\vec{\varphi}_0=\ln\left( \frac{A\,|\va\wedge\vb|^2}{4\vec{\beta}\cdot(\vec{\beta}-\va)}\right)$\\[7pt]
$\vec{\beta}\cdot\vec{\varphi}_0=\ln\left(  \frac{B\,|\va\wedge\vb|^2}{4\va\cdot(\va-\vb)}\right)$\\[-7pt]{}} \\
\hline
$P_2$ & $a_0\, t^{\frac{2}{\gamma_*^2}}$  & $\vec{\varphi}_{0}+\frac{2}{\gamma_*^2}\vg_*~\!\ln t$ & \makecell{\\[-7pt]$k=0$, $\vg_*=\vg_\perp$ \\[2pt]
$\vec{\alpha}\cdot\vec{\varphi}_0=\ln\left(\frac{\gamma_*^4A}{2(6-\gamma_*^2)}\frac{|\va-\vb|^2}{\vec{\beta}\cdot(\vec{\beta}-\va)}\right)$\\[7pt]
$\vec{\beta}\cdot\vec{\varphi}_0=\ln\left(\frac{\gamma_*^4B}{2(6-\gamma_*^2)}\frac{|\va-\vb|^2}{\va\cdot(\va-\vb)}\right)$\\
[-7pt]{}}  \\
\hline
\end{tabular}
\end{center}
\caption {The explicit solutions for $a(t)$ and $\vv(t)$ at the critical  points,  for the two-exponential potential \eqref{2exppot2} with 
  $\va\neq\vg_*\neq\vb$.~This is  the case where both exponentials scale in the same way at the critical point, see main text for more details.~For  
   $\vg_*=\va$ or $\vg_*=\vb$ the potential is dominated by one of the two exponentials, and  the solutions reduce to   those of Table~\ref{tab:fixedpointssols}, provided we 
replace $V_0$ therein by $A$ or $B$ respectively.~The condition  $V=0$  for $P_{\mathcal{C}}$, $P_0$ need only hold asymptotically.~The  
conditions   for $P_1$, $P_2$ admit a solution if and only if $\vg_\perp$, cf.\eqref{465},  lies within the allowed range for $\vg$, cf.~Fig.~\ref{fig8}.}
\label{tab:4}
\end{table}
\renewcommand{\arraystretch}{1}

\subsection{Stability}

The stability properties of the critical points for the two-exponential potential are determined by   the eigenvalues and  eigenvectors of  the five-dimensional 
system \eqref{system2},~\!\eqref{geom2}.~Let us first examine the case   $\va\neq\vg_*\neq\vb$.

$\bullet$ At $P_{\mathcal{C}}$ the   eigenvalues and eigenvectors of the single exponential potential, cf.~Table~\ref{tab:fixedpointsstab}, setting $\vg\rightarrow\vg_*$ therein, 
carry over to the  present case.~Imposing $\vn\cdot(\va-\vb)=0$, cf.~Table~\ref{tab:4}, there are no new eigenvalues that arise.~Two new eigenvectors, representing motion purely in the $\vg$-plane, are associated to 
the zero-eigenvalue.

$\bullet$   At  $P_0$ the eigenvalues and eigenvectors  of Table~\ref{tab:fixedpointsstab}, also extend  to the  present case.~In addition,  there is a double zero-eigenvalue associated with eigenvectors  purely in the $\vg$-plane.

$\bullet$ At $P_1$ the eigenvalues $-1\pm\frac{1}{\gamma_*}\sqrt{8-3\gamma_*^2}$ and the corresponding eigenvectors  of Table~\ref{tab:fixedpointsstab},  also extend here, with $\vg_*=\vg_\perp$, cf.~Table~\ref{tab:4}.~In addition there is a zero-eigenvalue with corresponding eigenvector whose projection on the $\vg$-plane is   perpendicular to $(\va-\vb)$.  This eigenvalue should  be discarded as unphysical, as it corresponds to motion that would take us outside the allowed range of $\vg$. 

Moreover, there are two eigenvalues given by,   
\eq{\label{435}
-1\pm\frac{1}{\gamma_*}\sqrt{\gamma_*^2-4~\!\va\cdot(\vg_*-\vb)}~,}
with corresponding eigenvectors whose projection on the $\vg$-plane is parallel to $(\va-\vb)$.  
These eigenvalues have negative real part if and only if $\va\cdot(\vg_*-\vb)>0$,  or equivalently $\vb\cdot(\vg_*-\va)>0$, where we took \eqref{411} into account. This will be true if we are in the case depicted in Fig.~\ref{fig4}, therefore we also must require  
$A,B>0$, otherwise $\vg_*$ is not  within the allowed range of $\vg$.  If instead  $\va\cdot(\vg_*-\vb)<0$,  we are in the 
case of Fig.~\ref{fig3b}, and one of the eigenvalues has positive real part.~We conclude that $P_1$, with $\vg_*=\vg_\perp$, is stable if and only if  $\theta_1,\theta_2<\frac{\pi}{2}$ and $A,B>0$.

The eigenvectors corresponding to \eqref{435} read,\footnote{Without loss of generality, in order to simplify the expressions, we have used the $O(2)$ symmetry of the theory, cf.~Footnote~\ref{f:4},  to impose $(\va-\vb)\propto(1,0)$ and, 
consequently, $\vg_\perp\propto(0,1)$. }
\eq{\label{436}
\left(-1\pm\frac{1}{\gamma_*}\sqrt{\gamma_*^2-4~\!\va\cdot(\vg_*-\vb)},0,0,-\sqrt{6}~\!\va\cdot(\vg_*-\vb),0\right)
~.}
The eigenvalue -2 of the  single exponential case is recovered in the $\va\rightarrow\vb$ limit, in which case we have $\vg_*=\va=\vb$, and the corresponding eigenvector reduces to the one of the single exponential potential.

$\bullet$ At $P_2$ the eigenvalues $\frac12(\gamma_*^2-6)$, $(\gamma_*^2-2)$ and the corresponding eigenvectors of Table~\ref{tab:fixedpointsstab} also extend here, with $\vg_*=\vg_\perp$, cf.~Table~\ref{tab:4}.~In addition there is a zero-eigenvalue  with corresponding eigenvector whose projection on the $\vg$-plane is perpendicular to $(\va-\vb)$, and  should therefore be discarded as unphysical. 

Moreover, there are two eigenvalues, 
\eq{
\frac14(\gamma_*^2-6)\left(1\pm\sqrt{1+\frac{8(\va\cdot\vb-\gamma_*^2)}{6-\gamma_*^2}}\right)~,}
with corresponding eigenvectors whose motion in the $\vg$-plane is parallel to $(\va-\vb)$.~The condition for the real part of both these eigenvalues to be negative is exactly as in the case discussed for $P_1$.~Therefore  $P_2$, with $\vg_*=\vg_\perp$,  is 
stable  if and only if $\gamma_*^2<2$ and $\theta_1,\theta_2<\frac{\pi}{2}$ and $A,B>0$.

We summarize the results of this analysis in Table~\ref{tab:5}.

\vfill\break

\renewcommand{\arraystretch}{2}
\begin{table}[H]
\begin{center}
\centering
\begin{tabular}{| c | c | c |}
\hline
\cellcolor[gray]{0.9} Point   &  \cellcolor[gray]{0.9} Eigenvalues &  \cellcolor[gray]{0.9} Stability  \\[4pt]
\hline
$P_{\mathcal{C}} $   & 4,\,0,\,$\sqrt{\frac32}(\sqrt{6}-\vg_*\cdot\vec{n})$  & unstable \\[7pt]
\hline
$P_0$   & $-2,\,1$,\,0  &  unstable  \\[7pt]
\hline
$P_1$   & \makecell{\\[-2pt]  $-2,\,-1\pm\frac{1}{\gamma_*}\sqrt{8-3\gamma_*^2}$,\,$-1\pm\frac{1}{\gamma_*}\sqrt{\gamma_*^2-4~\!\va\cdot(\vg_*-\vb)}$\\
[-7pt]{}}    &  \makecell{\\[-7pt] stable iff $\gamma^2_*>2$, \\ $\theta_1,\theta_2<\frac{\pi}{2}$ \& $A,B>0$ }\\[14pt]
\hline
$P_2$ & \makecell{\\[-7pt] $\frac12(\gamma_*^2-6)$, $\gamma_*^2-2$,\,$\frac14(\gamma_*^2-6)\left(1\pm\sqrt{1+\frac{8(\va\cdot\vb-\gamma_*^2)}{6-\gamma_*^2}}\right)$\\
[-7pt]{}}  &  \makecell{stable iff $\gamma^2_*<2$, \\  $\theta_1,\theta_2<\frac{\pi}{2}$ \& $A,B>0$ }\\[7pt]
\hline
\end{tabular}
\end{center}
\caption {Eigenvalues and stability of the critical points of the double exponential potential, for $\va\neq\vg_*\neq\vb$. 
 At the   stable critical point ($P_1$ if $\gamma_*^2>2$, or $P_2$ if $\gamma_*^2<2$) the effective exponent $\vg_*$ is given by $\vg_\perp$, cf.~\eqref{465} \& Table~\ref{tab:4},  and  we are in the case of Figure~\ref{fig4}. 
 }
\label{tab:5}
\end{table}
\renewcommand{\arraystretch}{1}

Next, let us  examine the case   $\vg_*=\va$;  we must also impose $A\geq0$, in order to ensure $V\geq0$ at the critical point.


$\bullet$ At $P_{\mathcal{C}}$,  in addition to  the   eigenvalues and eigenvectors of the single exponential potential, cf.~Table~\ref{tab:fixedpointsstab}, setting $\vg\rightarrow\vg_*=\va$ therein, 
we have one new eigenvalue, 
\eq{
\sqrt{6}~\!\vn\cdot(\va-\vb)~,}
corresponding to motion purely in the $\vg$-space.~One additional eigenvector corresponding to the zero eigenvalue 
represents motion along $\va$ and should be discarded as unphysical.

$\bullet$ At  $P_0$  the eigenvectors and eigenvalues are the same as for the case   $\va\neq\vg_*\neq\vb$ discussed previously.

\begin{figure}[H]
\begin{subfigure}{.5\textwidth}
  \centering
  \includegraphics[width=0.5\linewidth]{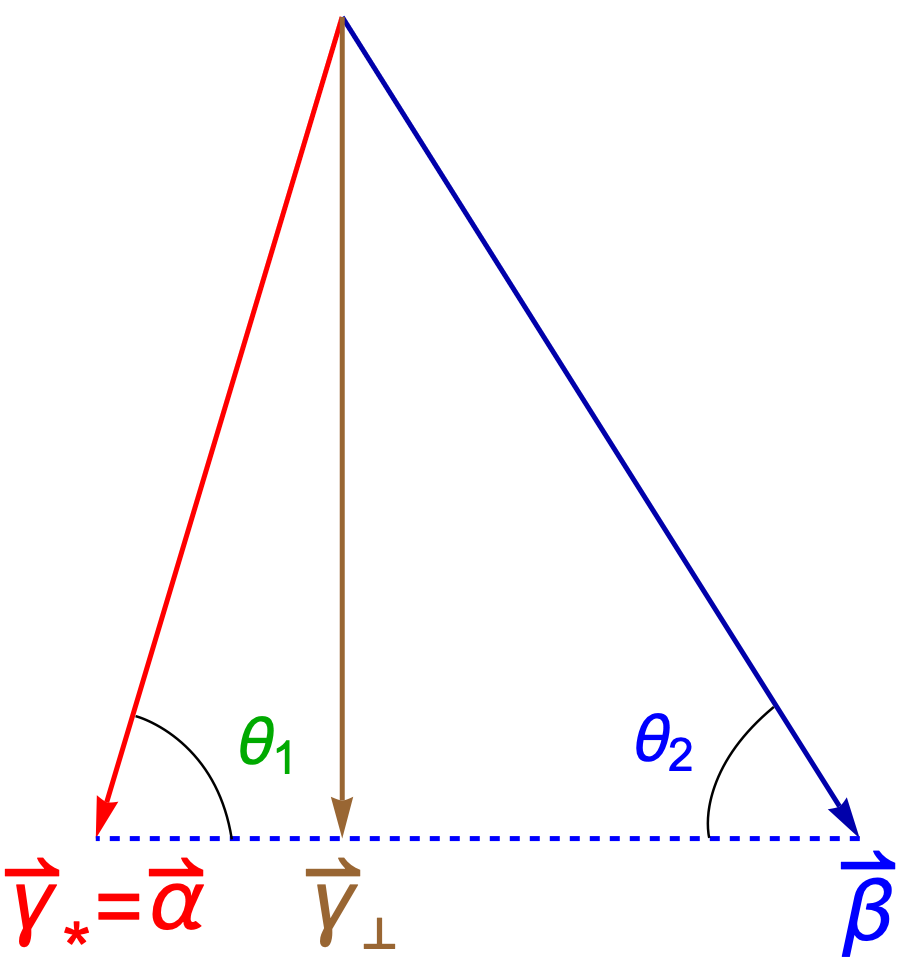}
  \caption{The case $\theta_1,\theta_2<\frac{\pi}{2}$: $\vg_*=\va$ is unstable;\\$\vg_\perp$ is in the  allowed range of $\vg$ if $A,B>0$.}
  \label{fig5a}
\end{subfigure}%
\begin{subfigure}{.5\textwidth}
  \centering
  \includegraphics[width=0.5\linewidth]{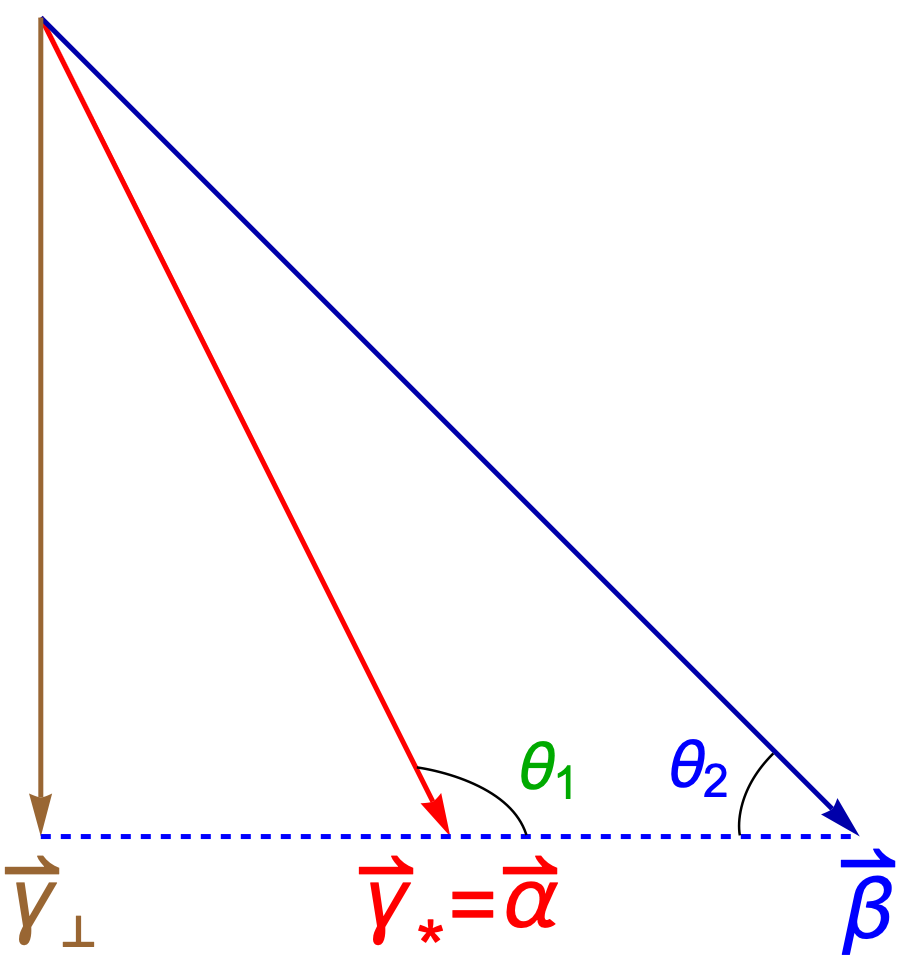}
  \caption{The case  $\theta_1>\frac{\pi}{2},\theta_2<\frac{\pi}{2}$:  $\vg_*=\va$ is stable;\\$\vg_\perp$ is in the  allowed range of $\vg$ if $A>0$, $B<0$.}
  \label{fig5b}
\end{subfigure}
\caption{$\vg_*=\va$ at the critical points $P_1$, $P_2$; we are assuming $A>0$ without loss of generality.}
\label{fig5}
\end{figure}

\begin{figure}[H]
\begin{subfigure}{.5\textwidth}
  \centering
  \includegraphics[width=0.5\linewidth]{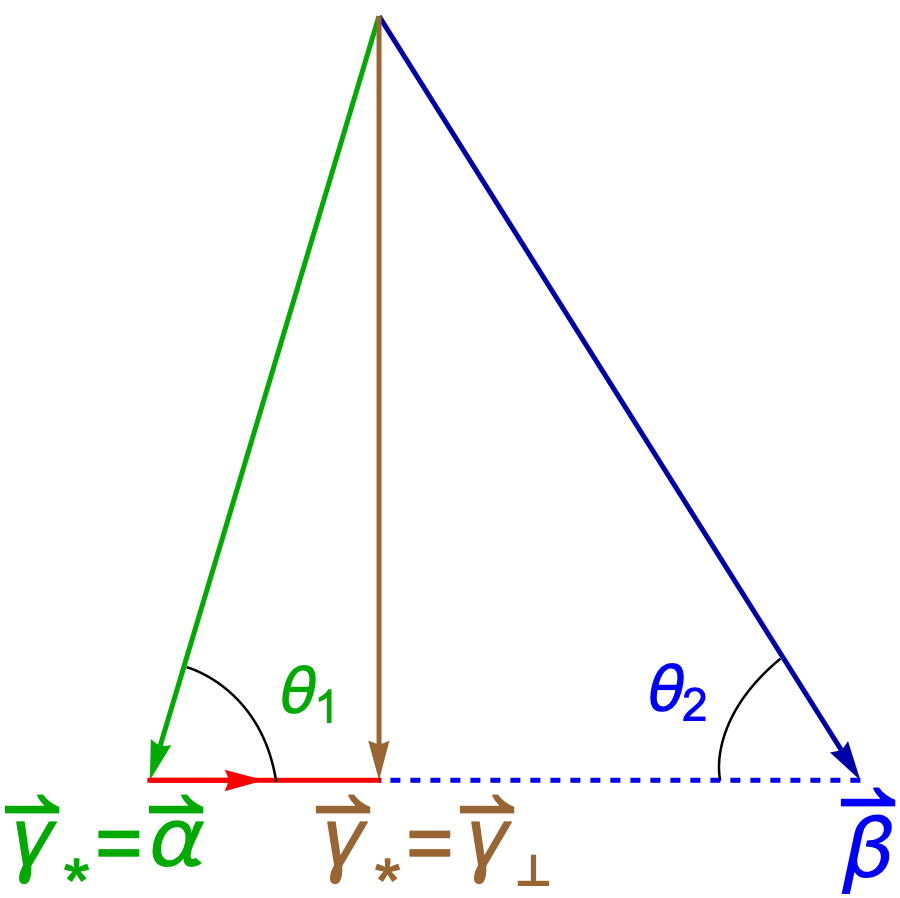}
  \caption{The case  $\theta_1,\theta_2<\frac{\pi}{2}$ \& $A,B>0$:\\ $\vg$ interpolates from $\vg_*=\va$ to $\vg_*=\vg_\perp$.}
  \label{fig6a}
\end{subfigure}%
\begin{subfigure}{.5\textwidth}
  \centering
  \includegraphics[width=0.5\linewidth]{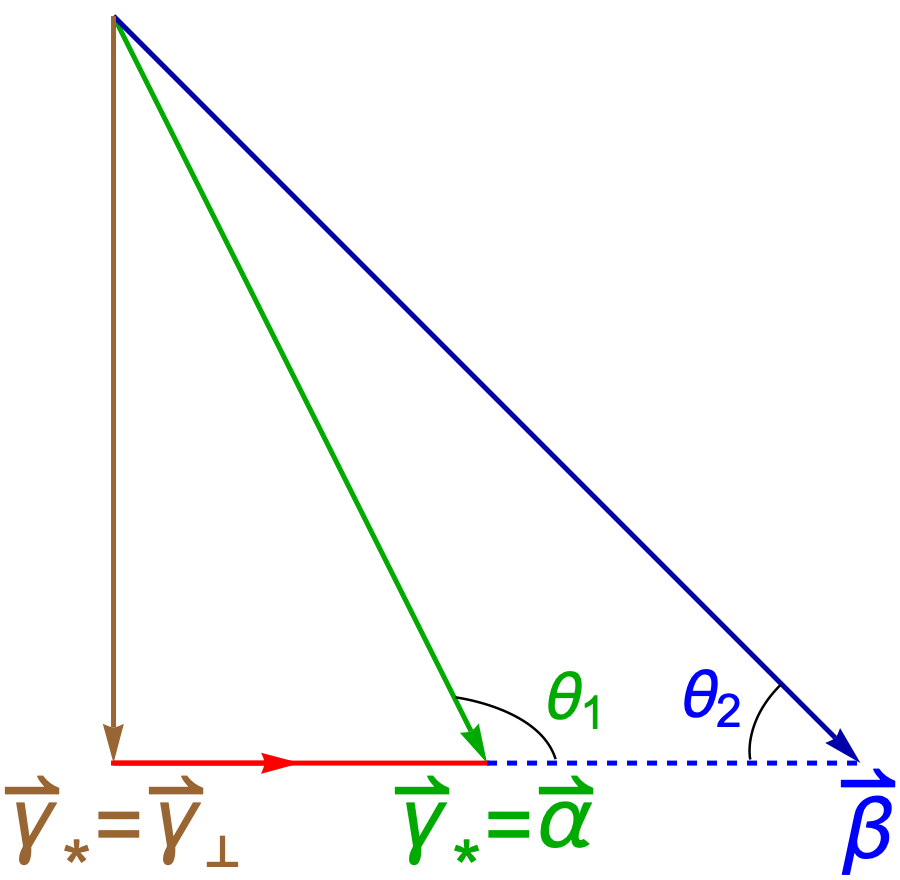}
  \caption{The case    $\theta_1>\frac{\pi}{2},\theta_2<\frac{\pi}{2}$ \& $A>0$, $B<0$:\\ $\vg$ interpolates from $\vg_*=\vg_\perp$ to $\vg_*=\va$.}
  \label{fig6b}
\end{subfigure}
\begin{subfigure}{.5\textwidth}
  \centering
  \includegraphics[width=0.5\linewidth]{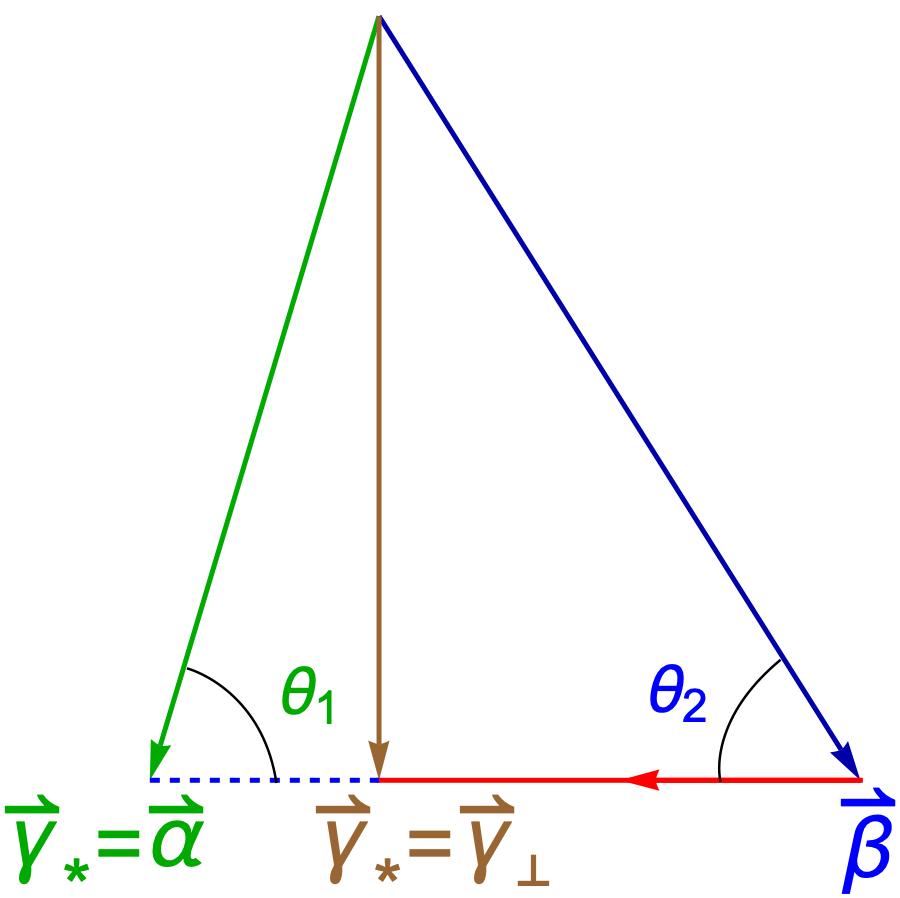}
  \caption{The case  $\theta_1,\theta_2<\frac{\pi}{2}$ \& $A,B>0$:\\ $\vg$ interpolates from $\vg_*=\vb$ to $\vg_*=\vg_\perp$.}
  \label{fig6c}
\end{subfigure}%
\begin{subfigure}{.5\textwidth}
  \centering
  \includegraphics[width=0.5\linewidth]{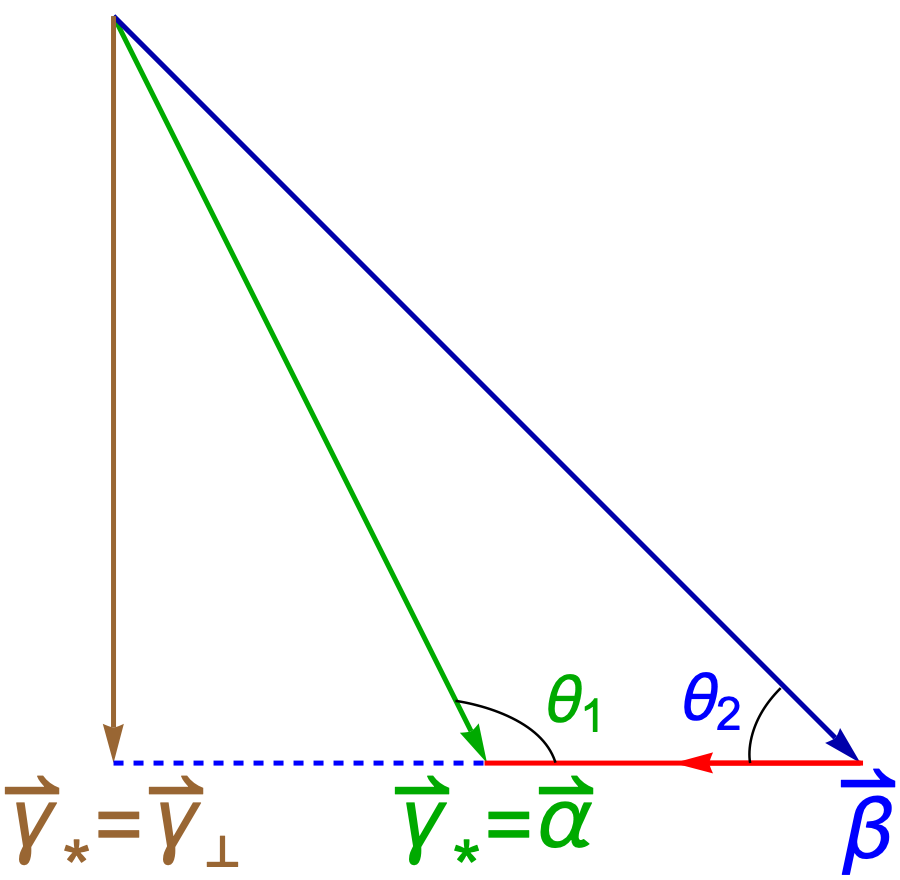}
  \caption{The case  $\theta_1>\frac{\pi}{2},\theta_2<\frac{\pi}{2}$ \& $A,B>0$:\\ $\vg$ interpolates from $\vg_*=\vb$ to $\vg_*=\va$.}
  \label{fig6d}
\end{subfigure}
\caption{ 
Interpolating flows in $\vg$-space, depicted with solid red lines,  from an unstable to a stable $\vg_*$; we are assuming $A>0$ without loss of generality. 
}
\label{fig6}
\end{figure}

$\bullet$   At  $P_1$,  in addition to  the   eigenvalues and eigenvectors of the single exponential potential,   there is an   unphysical zero-eigenvalue, with corresponding eigenvector whose projection on the $\vg$-plane is   perpendicular to $(\va-\vb)$. 
Moreover there is one new eigenvalue,
\eq{\label{evaqddf}\frac{2\va\cdot(\va-\vb)}{|\va|^2}~,}
with associated eigenvector corresponding to motion parallel to $(\va-\vb)$ in the $\vg$-plane. 
The eigenvalue \eqref{evaqddf} is positive if  $\theta_1<\frac{\pi}{2}$, and we are in the case of Fig.~\ref{fig5a}.~Otherwise,  \eqref{evaqddf} is negative and we are in the case of Fig.~\ref{fig5b}.~We conclude that 
$P_1$, with $\vg_*=\va$, is stable if and only if $\theta_1>\frac{\pi}{2}$. 
If in addition, $A>0$, $B<0$, then $\vg_\perp$ lies within the allowed range for $\vg$  and there can be solutions   interpolating in  $\vg$-space from the unstable $\vg_*=\vg_\perp$,  cf.~Fig.~\ref{fig3b}, to the 
  stable $\vg_*=\va$ cf.~Fig.~\ref{fig5b}. 
  If on the other hand  $A,B>0$, then $\vg_\perp$  lies within the allowed range for $\vg$,  provided $\theta_1<\frac{\pi}{2}$. In this case 
 there can be solutions  interpolating  in  $\vg$-space  from the unstable $\vg_*=\va$, cf.~Fig.~\ref{fig5a}, to the 
  stable $\vg_*=\vg_\perp$, cf.~Fig.~\ref{fig4}.  This discussion  extends  similarly to  the critical points with $\vg_*=\vb$. These results are  depicted in Fig.~\ref{fig6}.

%
%
%

%
%

$\bullet$   At  $P_2$,  in addition to  the   eigenvalues and eigenvectors of the single exponential potential,  
there is one new eigenvalue,
\eq{\label{evaqddf2}\va\cdot(\va-\vb)~,}
with associated eigenvector corresponding to motion parallel to $(\va-\vb)$ in the $\vg$-plane.~In 
addition there is an   unphysical zero-eigenvalue, with  eigenvector whose projection on the $\vg$-plane is   perpendicular to $(\va-\vb)$.~The eigenvalue \eqref{evaqddf2} is positive if and only if $\theta_1<\frac{\pi}{2}$.

We summarize the results of this analysis in Table~\ref{tab:6}.

\renewcommand{\arraystretch}{2}
\begin{table}[H]
\begin{center}
\centering
\begin{tabular}{| c | c | c |}
\hline
\cellcolor[gray]{0.9} Point   &  \cellcolor[gray]{0.9} Eigenvalues &  \cellcolor[gray]{0.9} Stability  \\[4pt]
\hline
$P_{\mathcal{C}} $   & 4,\,0,\,$\sqrt{\frac32}(\sqrt{6}-\va\cdot\vec{n})$,\,$\sqrt{6}~\!\vn\cdot(\va-\vb)$  & unstable \\[7pt]
\hline
$P_0$   & $-2,\,1$,\,0  &  unstable  \\[7pt]
\hline
$P_1$   & \makecell{\\[-2pt]  $-2,\,-1\pm\frac{1}{|\va|}\sqrt{8-3|\va|^2}$,\,$\frac{2}{|\va|^2}~\!\va\cdot(\va-\vb)$\\
[-7pt]{}}    &  \makecell{\\[-7pt] stable iff $|\va|^2>2$, \\  $\theta_1>\frac{\pi}{2},\theta_2<\frac{\pi}{2}$ \& $A>0$ }\\[14pt]
\hline
$P_2$ & \makecell{\\[-7pt] $\frac12(|\va|^2-6)$, $|\va|^2-2$,\,$\va\cdot(\va-\vb)$\\
[-7pt]{}}  &  \makecell{stable iff $|\va|^2<2$, \\   $\theta_1>\frac{\pi}{2},\theta_2<\frac{\pi}{2}$ \& $A>0$ }\\[7pt]
\hline
\end{tabular}
\end{center}
\caption {Eigenvalues and stability of the critical points of the double-exponential potential, for $\vg_*=\va$ and   $A\geq0$. 
 At the   stable critical point ($P_1$ if $|\va|^2>2$, or $P_2$ if $|\va|^2<2$)  we are in the case of Figure~\ref{fig5b}.~A 
 similar analysis can be performed for $\vg_*=\vb$, in which case we must have $B\geq0$.
 }
\label{tab:6}
\end{table}
\renewcommand{\arraystretch}{1}

\begin{figure}[H]
\begin{subfigure}{.5\textwidth}
  \centering
  \includegraphics[width=0.49\linewidth]{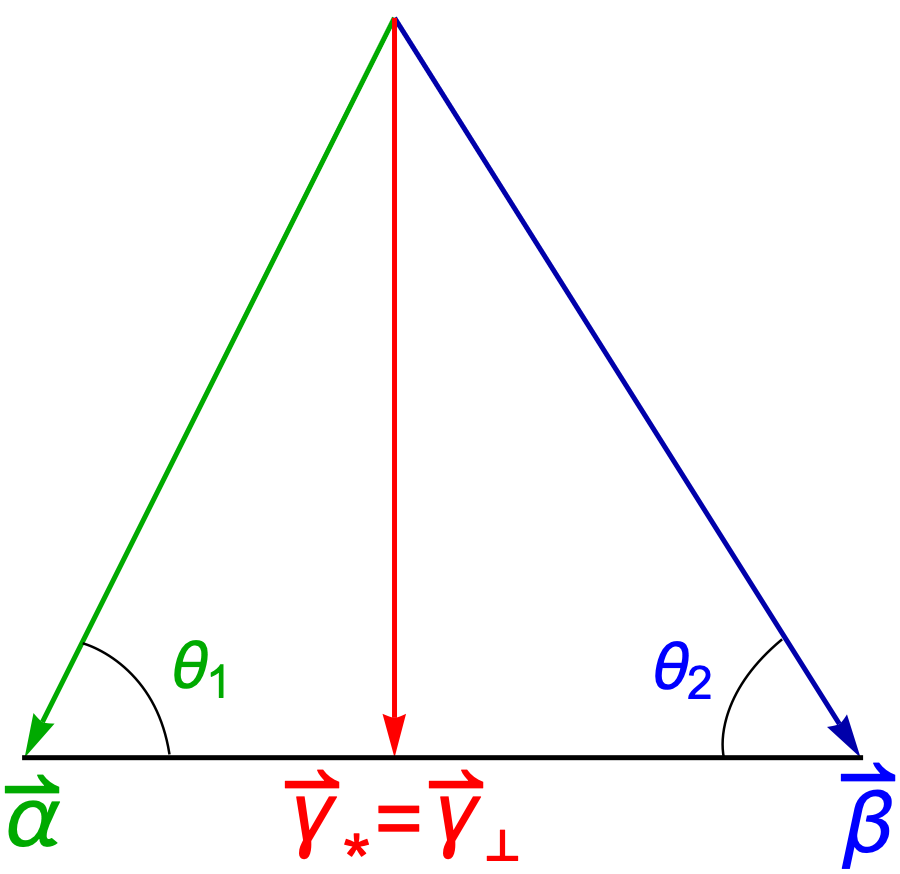}
  \caption{The case $\theta_1,\theta_2<\frac{\pi}{2}$ \& $A,B>0$: $\vg_*=\vg_\perp$.}
  \label{fig7a}
\end{subfigure}%
\begin{subfigure}{.5\textwidth}
  \centering
  \includegraphics[width=0.46\linewidth]{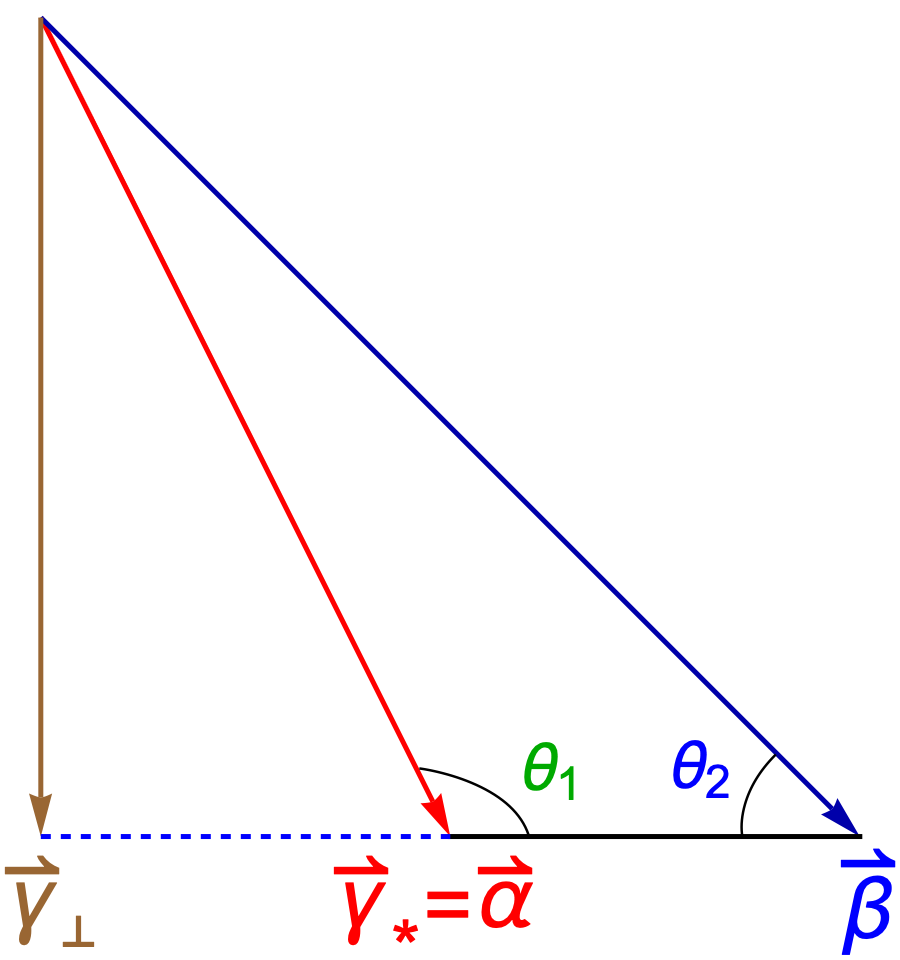}
  \caption{The case  $\theta_1>\frac{\pi}{2},\theta_2<\frac{\pi}{2}$ \& $A>0$:  $\vg_*=\va$.}
  \label{fig7b}
\end{subfigure}
\caption{$\vg_*$, depicted in $\vg$-space, at the stable critical point $P_1$ ($\gamma_*^2>2$), or $P_2$ ($\gamma_*^2<2$); we are assuming $A>0$ without loss of generality.~In 
all cases,  $\vg_*=\vg_{ch}$, where   $\vg_{ch}$  is the shortest vector connecting the origin   to the convex hull of  $\va$, $\vb$ (solid black line).~Conversely, $P_{1,2}$ are unstable if $\vg_{ch}$ lies outside the allowed range of $\vg$.}
\label{fig7}
\end{figure}

 In all cases, at the fully stable critical point ($P_1$ if $\gamma_*^2>2$ or $P_2$ if $\gamma_*^2<2$), $\gamma_*$ is equal to the distance of the origin of $\vg$-space to the {\it convex hull} of the exponents $\va$, $\vb$, the interval depicted in black in Fig.~\ref{fig7}.~I.e.~let $\vg_{ch}$ be the shortest vector connecting the origin of $\vg$-space to the convex hull of $\va$, $\vb$;~at the fully stable critical point we have,
\eq{ \vg_*=\vg_{ch}~.}
Conversely, if $\vg_{ch}$ lies outside the allowed range of $\vg$, then $P_{1,2}$ are unstable.

 Since here we are only dealing with 
  two exponentials, introducing the notion of the convex hull is somewhat redundant: it is simply  the base of the triangle formed by the two vectors  $\va$, $\vb$.  However, it becomes useful in the context of  multi-exponential potentials, see e.g.~\cite{Shiu:2023fhb}.  Note that the distance of the origin to the convex hull of  $\va$, $\vb$ is not always the same as the distance to the line defined by  $\va$, $\vb$. Indeed,  $\gamma_*=|\vg_\perp|$ in the case of Fig.~\ref{fig7a}, whereas $\gamma_*>|\vg_\perp|$ in the case of Fig.~\ref{fig7b}.

\vfill\break

%
%

\section{Universal compactifications}\label{sec:universal}

In \cite{Marconnet:2022fmx} we established a  cosmological consistent truncation of IIA supergravity to a 4d gravitational theory coupled to two scalar fields of the form \eqref{2sc}, originating from the 10d dilaton and the  warp factor of the metric.~The term ``cosmological consistent truncation''  means a repackaging of the 10d equations of motion, such that every FLRW  solution of the 4d theory lifts to a  solution of 10d IIA supergravity.\footnote{All solutions presented in  \cite{Marconnet:2022fmx} 
 can be thought of as solutions of an appropriate 1d consistent truncation, of the kind  used in minisuperspace.~However, not all of the solutions of  \cite{Marconnet:2022fmx}  admit a 4d consistent truncation.  This is reflected in the fact that the potential of the 1d truncation, 
 given in  \cite[Eq.~(15)]{Marconnet:2022fmx}, is much richer than the 4d one, given here  in \eqref{4dpot}.} 
  Explicitly, the 4d potential is given by,\footnote{\label{foot:14}Unlike  in \cite{Marconnet:2022fmx}, we have canonically normalized the scalar fields.  Comparison of \eqref{2sc} of the present paper with the action of \cite[Eq.~(22)]{Marconnet:2022fmx} gives, 
$A^{\text{there}}=\sqrt{\frac{\pi G}{3}}~\!\varphi_1$, 
$\phi^{\text{there}}=\sqrt{ 16\pi G}~\!\varphi_2$, 
$V^{\text{there}}= 16\pi G ~\!V$. 
} 
\eq{\spl{\label{4dpot}
V=\left\{
\begin{array}{ll}36b_0^2   e^{-\sqrt{6} \varphi_1-\sqrt{2} \varphi_2}
+\frac34c_0^2  e^{ -\frac{7\varphi_1}{\sqrt{6}}  +\frac{\varphi_2}{\sqrt{2}}  }  &~  \text{\small{CY~internal~3-~and~4-form~fluxes}} \\ \\
\frac14 c_{\varphi}^2 e^{-3 \sqrt{\frac{3}{2}} \varphi_1-\frac{\varphi_2}{\sqrt{2}}}+\frac14m^2 e^{-\sqrt{\frac{3}{2}}\varphi_1+\frac{5 \varphi_2}{\sqrt{2}}} 
-3 \lambda  e^{-2 \sqrt{\frac23}\varphi_1} &~  \text{\small{E~external~4-form~flux}} \\ \\
\frac34c_0^2 e^{-\frac{7 \varphi_1}{\sqrt{6}}+\frac{\varphi_2}{\sqrt{2}}}
+\frac14m^2 e^{-\sqrt{\frac{3}{2}}\varphi_1+\frac{5 \varphi_2}{\sqrt{2}}} 
-3 \lambda  e^{-2 \sqrt{\frac23}\varphi_1} &~  \text{\small{EK~internal~4-form~flux}} \\ \\
\frac14 c_{\varphi}^2 e^{-3 \sqrt{\frac{3}{2}} \varphi_1-\frac{\varphi_2}{\sqrt{2}}}+\frac34c_f^2 
e^{-\frac{5 \varphi_1}{\sqrt{6}}+\frac{3 \varphi_2}{\sqrt{2}}}
-3 \lambda  e^{-2 \sqrt{\frac23}\varphi_1}  &~  \text{\small{EK~internal~2-form, external~4-form}}
 \end{array} \right.
}}
where E, EK, or CY stands for a 6d compactification manifold of Einstein, Einstein-K\"{a}hler, or Calabi-Yau type. 
Depending on the class of the compactification manifold, and  in order to preserve the consistency of the 4d truncation, only certain  types of   fluxes can be simultaneously  turned on in the 10d theory. 
The remnant of these 10d fluxes is manifested in the potential of the 4d theory in the presence of certain constants, listed below in Table~\ref{tab:constants}.~These are precisely the constants of proportionality between the 10d fluxes and the universal forms of the compactification (volume,  K\"{a}hler form, etc), and will  in general be subject to quantization in the  full-fledged quantum theory. 
\begin{table}[H]
\begin{center}
\begin{tabular}{|c|c|}
  \hline
   $m$ & zero-form (Romans mass) \\
    \hline
   $c_f$ & internal two-form  \\
     \hline
$b_0$ &internal  three-form \\
  \hline  
$c_\varphi$ & external four-form \\
  \hline
$c_0$ &internal  four-form \\
  \hline
$\lambda$ & scalar curvature of $M_6$\\
  \hline
\end{tabular}
\end{center}
\caption{List of the constant coefficients appearing  in the potential \eqref{4dpot} of the 4d consistent truncation, and their 10d origin.~A form is called external (internal), if all its legs are along the 4d external (6d internal) directions.}
\label{tab:constants}
\end{table}
%
%

Two-exponential models, with potential as in \eqref{2exppot2}, can be formed by selecting two species of flux from within each class of compactification — corresponding to each  of the four lines on the right-hand side of \eqref{4dpot} — and reading off the corresponding constants $A$,$B$ and  exponents $\va$, $\vb$.~The  resulting models are listed in 
Table \ref{tab:modelscan}.~At least one of  $A$, $B$ is  always  positive, since the only potentially negative coefficient in the potential  \eqref{4dpot}  is $-3\lambda$, coming from compactification on a 6d Einstein space with scalar curvature $\lambda$.~According to our convention we will always take $A>0$, and so we will set  $B=-3\lambda$ in all models coming from compactification on 6d Einstein spaces.~This choice allows for $B$ both positive or negative, corresponding to  negative or positive 6d curvature respectively.~For all models  the allowed range of $\vg$ is   as  in Fig.~\ref{fig8}.

For all the models in Table \ref{tab:modelscan}, we have $\gamma_*^2>2$, in accordance with the swampland conjectures \cite{Obied:2018sgi,Hebecker:2018vxz,Andriot:2019wrs,Lust:2019zwm,Bedroya:2019snp,Andriot:2020lea,Rudelius:2021oaz,Rudelius:2021azq}.~This implies that the  fully stable critical point is always $P_1$,  requiring an open universe. 
For the models with  a 6d curvature contribution, i.e.~those with $B=-3\lambda$, we must require $\lambda<0$ for $\vg_{ch}$ to lie in the allowed range for $\vg$, so that $P_1$ is stable.

The smallest value  for the effective exponent at the critical point is  attained for the model from compactification on an Einstein space with negative curvature and non-vanishing Romans mass.  For this model we have $\gamma_*^2=\frac{50}{19}$, which  is smaller than the threshold 
value, Eq.~\eqref{scase}, below which $P_1$ is a stable node \cite{Andriot:2023wvg}.~To our knowledge, this is the smallest effective exponent so far obtained from a universal compactification model.

Five of the  models of Table \ref{tab:modelscan} correspond to the case depicted in Fig.~\ref{fig4}.  The remaining three models have 
 $\vg=\vb$ at the critical point, corresponding  to the case depicted in Fig.~\ref{fig5b}, with  ($A$, $\va$, $\theta_1$) and 
  ($B$, $\vb$, $\theta_2$) interchanged.  In  these models  $\vg_\perp$ is always outside the allowed range for $\vg$, so there are no possible interpolating flows of the type depicted in Fig.~\ref{fig6b}.

\subsection{Late-time physics}

As can be seen from \cite[Eqs.~\!(1),(4)]{Marconnet:2022fmx} and Footnote \ref{foot:14}, setting  $8\pi G=1$, 
the uplift of the  universal cosmologies to the 10d Einstein frame is given by,
\eq{\label{5.2}
\d s_{10}^2=e^{-\sqrt{\frac32}\varphi_1}\d s^2+e^{\frac{1}{\sqrt{6}}\varphi_1}\d s^2(M_6)~;~~~e^\phi=e^{\sqrt{2}\varphi_2}
~,}
where $\phi$ is  the 10d dilaton, $\d s^2$ is the FLRW metric \eqref{metric4} in the 4d Einstein frame,  and $\d s^2(M_6)$ is the time-independent metric of the 
internal compactification manifold $M_6$.  

In all cases, the late-time asymptotics ($t\rightarrow\infty$) are dictated by the attractor point $P_1$, and the associated solution, cf.~Table~\ref{tab:4}. This implies  that at late times the solution behaves as, 
\eq{\label{5.3}
a(t)= \frac{~~\gamma_*}{\sqrt{\gamma_*^2-2}}\, t +\mathcal{O}(t^{-1})~;~~~\vv= \vv_{0}+\frac{2}{\gamma_*^2}\vg_*~\!\ln t+\mathcal{O}(t^{-1})
~.
}
In particular, we will be  interested in the  late-time behavior of the string coupling and the Hubble length,  
\eq{ g_s:=e^\phi~, ~~~L_H:=\frac{1}{H}~,}
as well as the effective size of the compactification  manifold.~The latter is defined as the time-dependent  Kaluza-Klein (KK) scale of $M_6$, as measured in the 4d Einstein frame; it can be read off of 
the 10d metric \eqref{5.2}, 
\eq{\label{5.5}
L_6(t):=L_6 e^{\sqrt{\frac23}\varphi_1}~,}
where $L_6$ is the KK scale of $M_6$ in the absence of a warp factor, see \cite{Andriot:2025cyi} for a recent discussion. Taking Eqs.~\eqref{5.2}, \eqref{5.3} into account,  we thus have, 
\eq{\label{5.6}
g_s\rightarrow  e^{\sqrt{2}(\vv_{0})_2}~t^{2\sqrt{2}~\!\frac{(\vg_{*})_2}{\gamma_*^2}}  ~;~~~L_H\rightarrow t~;~~~L_6(t)\rightarrow  L_6~\! e^{\sqrt{\frac23}(\vv_{0})_1} ~ t^{2\sqrt{\frac23}~\!\frac{(\vg_{*})_1}{\gamma_*^2}}
~,
}
in the $t\rightarrow\infty$ limit.

We will  now briefly review the derivation of Eq.~\eqref{5.5}, in a way that will also 
allow us to obtain an estimate of higher-order derivative $\alpha'$-corrections.  
For our purposes it is sufficient to consider the following 10d effective action,
\eq{
S_{10}=\int\d^{10}x\sqrt{-g_{10}}~\! \sum_{n=0}^\infty (\alpha')^n \left(R_{10}\right)^{n+1}
~,}
where $R_{10}$,  $g_{10}$ is the scalar curvature,  the determinant of the metric \eqref{5.2}, respectively.~Ignoring derivatives of $\varphi_1$, we can rewrite the above as, 
\eq{
S_{10}=\int\d^{4}x\sqrt{-g_{4}} \int\d^{6}x\sqrt{g_{6}}~\! \sum_{n=0}^\infty \left(\alpha' e^{\sqrt{\frac32}\varphi_1}\right)^n \left(R_{4}+e^{-2\sqrt{\frac23}\varphi_1}R_6\right)^{n+1}
~,}
where $g_6$, $R_6$ is the determinant,  scalar curvature of $\d s^2(M_6)$;  $g_{4}$, $R_4$ is the determinant,  scalar curvature of $\d s^2$, cf.~Eq.~\eqref{5.2}. From this, and the fact that $R_6$ scales as $1/L_6^2$,  
we can see that, from the point of view of the 4d Einstein frame, the 6d effective scale is given by $L_6(t)$ of Eq.~\eqref{5.5}. In addition, we see that for the $\alpha'$-corrections to be small, we must have,
\eq{
 \alpha' e^{\sqrt{\frac32}\varphi_1} R_{4} \ll 1~~~\text{\&}~~~
  \alpha'   e^{-\frac{1}{\sqrt{6}}\varphi_1}R_6
  \ll1
~.}
Given that $R_6$ scales as $1/L_6^2$, while  $R_4$ scales as $1/L_H^2$ \cite{Andriot:2025cyi}, we arrive at the condition, 
\eq{
\text{Small~$\alpha'$-corrections}~~~ \Leftrightarrow ~~~\frac{L_H}{L_6(t)}~\! e^{\frac{1}{2\sqrt{6}}\varphi_1}\gg \frac{l_s}{L_6}~~~\text{\&}~~~e^{\frac{1}{2\sqrt{6}}\varphi_1} \gg \frac{l_s}{L_6}
~,}
where $l_s:=\sqrt{\alpha'}$ is the string length.~In the absence of warp factor ($\varphi_1=0$), the last inequality reduces to the condition 
that the size of the compactification space should be large in string units — which is the usual  necessary condition for the validity of the supergravity description of time-independent solutions. 

Furthermore, 
let us define the {\it absence of decompactification}, as the condition that the effective KK length of the internal space should be at most of the order of the Hubble length, 
\eq{\label{ratio1}
 \text{Absence~of~decompactification} ~~~ \Leftrightarrow ~~~\frac{L_H}{L_6(t)}\gtrsim 1~.}
This condition prevents the internal space from expanding at a faster rate than the 4d spacetime, but  
allows for backgrounds for which the two scales are of the same order, analogous to AdS vacua of Freund-Rubin type. 
Moreover it follows that,
\eq{
 \text{Absence~of~decompactification~~~\&}~~~ e^{\frac{1}{2\sqrt{6}}\varphi_1} \gg \frac{l_s}{L_6}~~~\Rightarrow ~~~\text{Small~$\alpha'$-corrections} 
}
Absence of decompactification is weaker than the condition of {\it scale separation}, which is usually defined as the condition that the radius of curvature   of the 4d space should be much greater than the KK length of the internal space —a necessary condition for the  10d theory to admit a 4d low-energy effective description. Adapting this definition to the case of our time-dependent backgrounds, we have,
\eq{\label{ratio2}
\text{Scale~separation} ~~~ \Leftrightarrow ~~~ \frac{L_H}{L_6(t)}\gg 1~.}
For time-independent supergravity  backgrounds originating from string/M-theory, the condition of scale separation is notoriously difficult to satisfy with only classical ingredients,  in the absence of orientifolds   \cite{Tsimpis:2012tu, Richard:2014qsa, Marchesano:2020qvg, Gautason:2015tig, Font:2019uva, Lust:2020npd, Junghans:2020acz, Farakos:2020phe, DeLuca:2021mcj, Cribiori:2021djm, Tsimpis:2022orc, Apers:2022zjx, Farakos:2023wps,  Coudarchet:2023mfs, Arboleya:2024vnp, Cribiori:2024jwq, VanHemelryck:2025qok,Tringas:2025uyg}.\footnote{It was recently shown that scale separation is possible within the 6d Salam-Sezgin model \cite{Proust:2025vmv}.} 
However, as was noted in \cite{Shiu:2023fhb, Andriot:2023wvg}, and  recently emphasized in \cite{Andriot:2025cyi}, the situation is different for 
  cosmological backgrounds.~Besides the first model of Table~\ref{tab:modelscan}, for which higher-order corrections blow up at late times and the internal space decompactifies, we distinguish the following cases.

{\it Models with scale separation}:~the three models with  $\gamma_*^2=8$. They all  
have the same effective exponent $\vg_*=\vg_\perp$, due to the fact that 
 the potential in these models consists of exponential terms — those generated by the fluxes $c_f$, $c_\varphi$, $m$, $c_0$ — 
whose exponents all lie on the same line, cf.~Table~\ref{tab:modelscan}. 

{\it Models with curvature domination}:~the three models with $\vg_*=\vb=(2\sqrt{\frac23},0)$, whose potential is dominated at future infinity by the exponential associated 
with the  internal curvature $\lambda$. These have $g_s\rightarrow\text{cnst}$, $L_6(t)\rightarrow t$ asymptotically, so they all exhibit absence of decompactification but no time-evolution driven  scale separation.~More 
precisely, the asymptotic behavior near the future attractor $P_1$ can be read off of Table~\ref{tab:fixedpointssols} and  Eq.~\eqref{5.6}, 
\eq{\label{514}
g_s\rightarrow  e^{\sqrt{2}(\vv_{0})_2}~;~~~
\frac{L_6(t)}{L_H}\rightarrow  L_6~\!\sqrt{2|\lambda|}~.}
Moreover, the $(\vv_0)_2$ component remains undetermined asymptotically, so that $g_s$ is a modulus, within the supergravity approximation,  and can be tuned to be as small as desired. 

On the other hand, 
the ratio $L_6(t)/L_H$ cannot be tuned by scaling the time-independent overall length $L_6$ of $M_6$, since the scalar curvature $\lambda$ itself scales 
as $1/L_6^2$.~It might  be possible, however, to obtain some degree of  {\it relative}  (as opposed to {\it parametric}) scale separation by modding out $M_6$ by discrete subgroups of its isometry group, as in \cite{Collins:2022nux,Tsimpis:2022orc}.

\begin{figure}[H]
\begin{center}
\includegraphics[width=.3\textwidth]{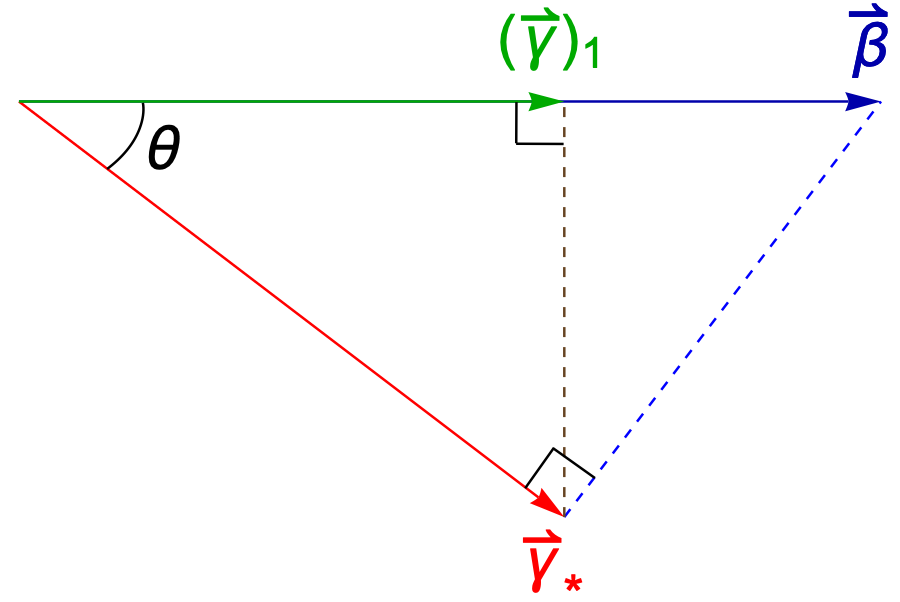}
\caption{Model with 6d curvature term, in the case where the latter contributes but does not  dominate asymptotically, i.e.~$\vg_*=\vg_\perp\neq\vb$. The 
relation ${(\vg_{*})_1}/{\gamma_*}={\gamma_*}/{|\vb|}=\cos\theta$, ensures the same asymptotic scaling for $L_6(t)$, as in the case of curvature domination.}\label{curvature}
\end{center}
\end{figure}

{\it Model with curvature contribution}:~fourth model of Table~\ref{tab:modelscan}.~Its potential includes an exponential term proportional to $\lambda$, which 
  contributes non-trivially asymptotically, but does not dominate (so 
that $\vg_*=\vg_\perp\neq\vb$). 
In this case we have the exact same scaling for $L_6(t)$ asymptotically, as for the models with curvature domination.~This follows from Eq.~\eqref{5.6} and the fact that  ${(\vg_{*})_1}/{\gamma_*^2}={1}/{|\vb|}$. 
The latter   has a simple geometric interpretation, cf.~Fig.~\ref{curvature}. 

We conclude that models with asymptotically nontrivial curvature contribution, whether or not the latter dominates the potential, cannot achieve time-evolution driven scale separation. 
Instead, these models exhibit absence of decompactification, whereby the FLRW factor scales at late times in the same way as the internal space.~This feature is independent of 
 the ratio  $A/B$.~In other words, the asymptotic scaling of $L_6(t)$ is 
independent of the strength of the contribution of the curvature term to the potential, as long as this contribution 
is non-vanishing asymptotically.~This generalizes the observations of \cite{Andriot:2025cyi} for curvature dominated potentials, to potentials with merely curvature contribution, in line with the results of  \cite{Shiu:2023fhb} for flat universe.

On the other hand,  the precise coefficient of the ratio of internal-to-external length  depends on the exponents of the potential.~For example, for the model at hand, this ratio can be calculated from Eq.~\eqref{5.6}, using the condition on $\vv_0$ from Table~\ref{tab:4}, 
\eq{\label{524}
\frac{L_6(t)}{L_H}\rightarrow \frac{5}{2\sqrt{3}}~\! L_6~\!  \sqrt{|\lambda|}~,}
which is different from   \eqref{514}.~In particular, the numerical coefficient of the ratio depends on the angle of $(\va,\vb)$, 
\eq{\label{euh}
\frac{L_6(t)}{L_H}\rightarrow      L_6~\!   |\va\wedge\vb| ~\!  \sqrt{\frac{3|\lambda|}{4\va\cdot(\va-\vb)}}     ~,
}
and it  tends to zero as  this angle tends to $\pi$.~It  thus might be possible, by tuning this angle, 
 to achieve an amount of relative scale separation, in models with curvature contribution.

\vfill\break
 
 
\renewcommand{\arraystretch}{2}
\begin{table}[H]
\begin{center}
\centering
\begin{tabular}{| c | c | c | c | c | c | c | c | c | c | c | c |}
\hline
\cellcolor[gray]{0.9}$A$ &\cellcolor[gray]{0.9}$B$ &  \cellcolor[gray]{0.9} $\color{green}\va$, $\color{blue}\vb$, $\color{red}\vg_\perp$ &   \cellcolor[gray]{0.9} $\vg_\perp$ &  
\cellcolor[gray]{0.9}     $\vg_*$  & 
\cellcolor[gray]{0.9} $\gamma_*^2$  & 
\cellcolor[gray]{0.9} $g_s$ & 
\cellcolor[gray]{0.9} $e^{\sqrt{\frac{2}{3}}\varphi_1}$ & 
\cellcolor[gray]{0.9} $\frac{L_6(t)}{L_H}$  &
\cellcolor[gray]{0.9} $g_s,~\!\alpha'$ & 
\cellcolor[gray]{0.9} Sc.~\!S. \\[4pt]
\hline
$\frac34 c^2_0$ &  $36b^2_0$ & \begin{minipage}[c][2cm][c]{.15\textwidth}
\centering
      \includegraphics[width=20mm]{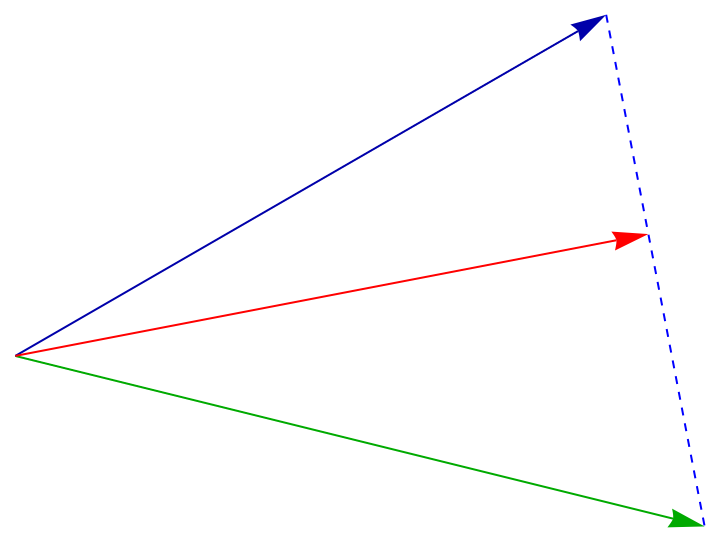}
    \end{minipage}   & Y & $\vg_\perp$ &  $\frac{50}{7}$  & $t^{\frac15}$
  &   $t^{\frac35}$ & $t^{\frac25}$ & N & N \\[7pt]
\hline
$\frac14c_\varphi^2$ & $\frac14m^2$ & \begin{minipage}[c][2cm][c]{.15\textwidth}
\centering
      \includegraphics[height=18mm]{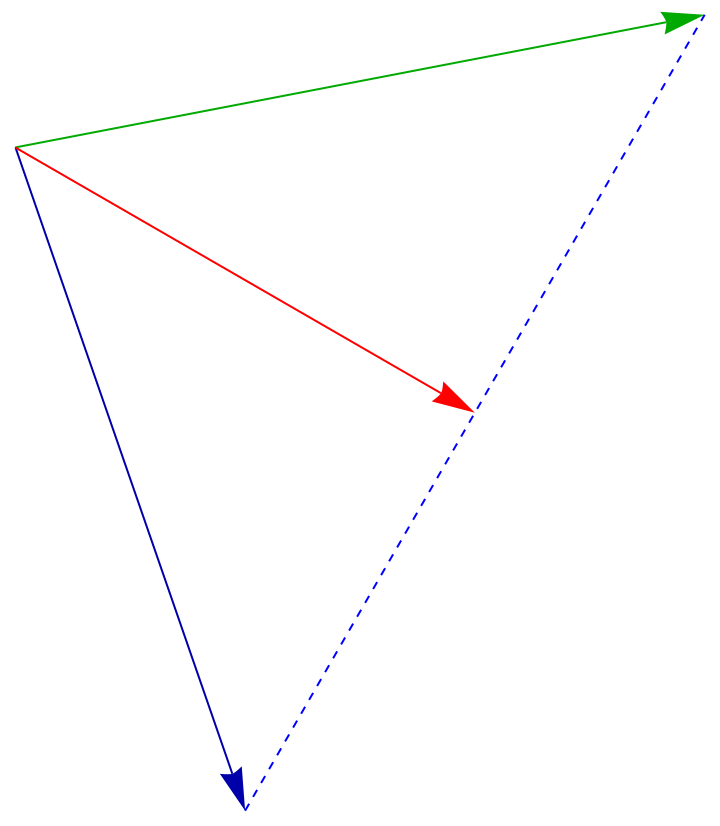}
    \end{minipage}     & Y  & $\vg_\perp$ &  $8$  & $t^{-\frac12}$&$t^{\frac12}$ &$t^{-\frac12}$ & Y&Y \\[7pt]
 \hline
$\frac14c_\varphi^2$ & $-3\lambda$ & \begin{minipage}[c][2cm][c]{.15\textwidth}
\centering
      \includegraphics[ width=23mm ]{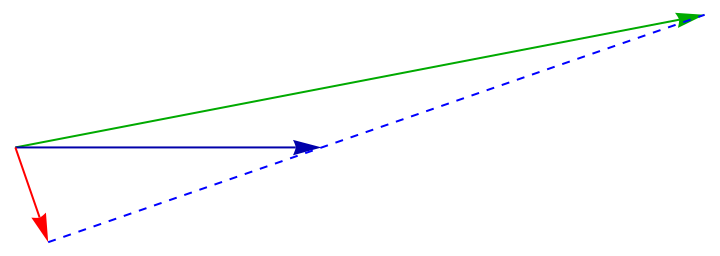}
    \end{minipage}     & N & $\vb$ & $\frac83$    & $t^{0}$&$t^1$ &$t^{0}$ & Y& R\\[7pt]
 \hline
$\frac14m^2$ & $-3\lambda$ & \begin{minipage}[c][2cm][c]{.15\textwidth}
\centering
      \includegraphics[height=19mm]{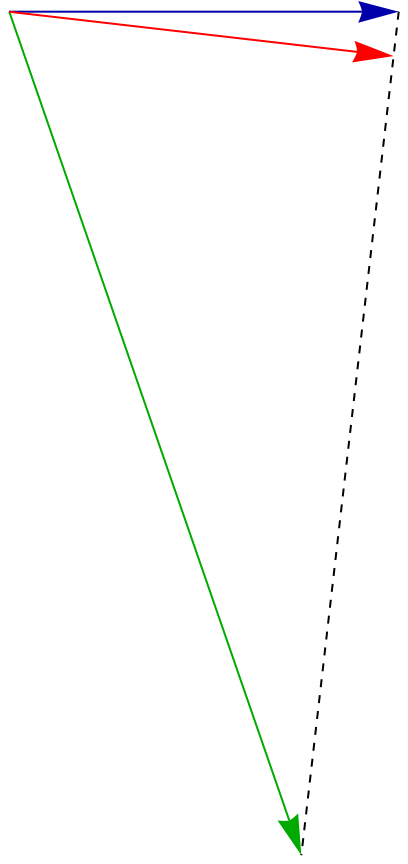}
    \end{minipage}     & Y & $\vg_\perp$ & $\frac{50}{19}$    & $t^{-\frac15}$&$t^1$ &$t^{0}$ & Y & R\\[7pt]
 \hline
$\frac34c_0^2$ & $\frac14m^2$ & \begin{minipage}[c][2cm][c]{.15\textwidth}
\centering
      \includegraphics[height=19mm]{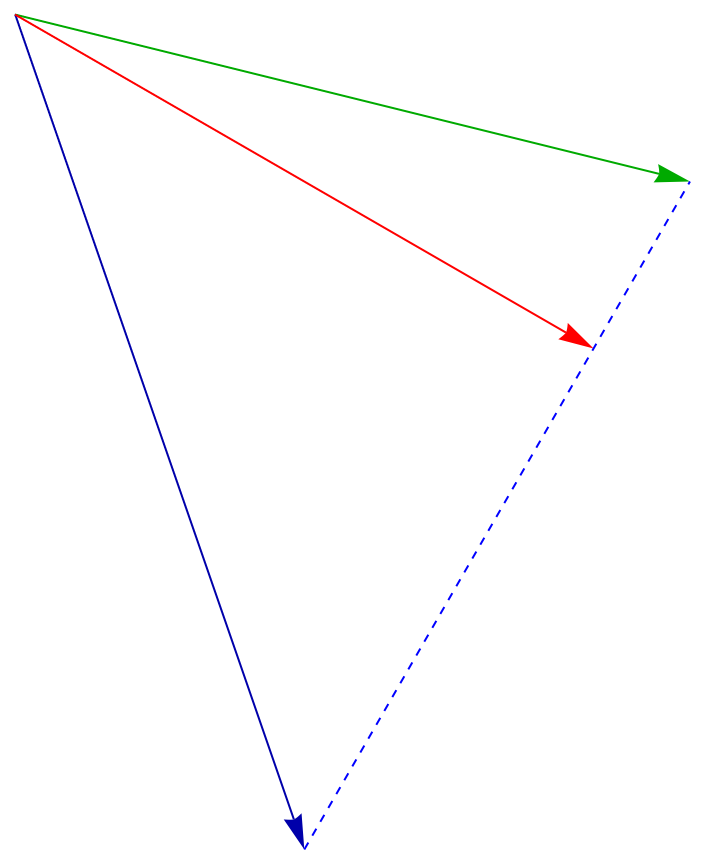}
    \end{minipage}    & Y & $\vg_\perp$ & $8$   & $t^{-\frac12}$&$t^{\frac12}$ &$t^{-\frac12}$  & Y& Y\\[7pt]
  \hline
$\frac34c_0^2$ & $-3\lambda$ & \begin{minipage}[c][2cm][c]{.15\textwidth}
\centering
      \includegraphics[width=23mm]{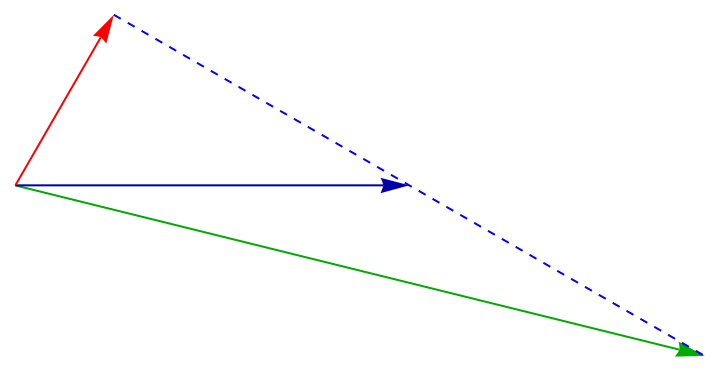}
    \end{minipage}     & N & $\vb$ &  $\frac83$    & $t^{0}$&$t^1$ &$t^{0}$ & Y & R\\[7pt]
   \hline
$\frac14c^2_\varphi$ & $\frac34c_f^2$ & \begin{minipage}[c][2cm][c]{.15\textwidth}
\centering
      \includegraphics[width=23mm]{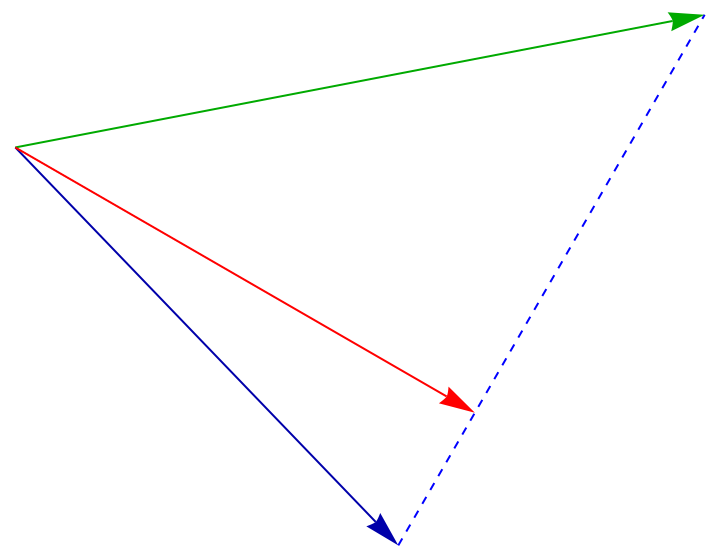}
    \end{minipage}     & Y & $\vg_\perp$ &  $8$    & $t^{-\frac12}$&$t^{\frac12}$ &$t^{-\frac12}$ & Y& Y\\[7pt]
   \hline
$\frac34c_f^2$ & $-3\lambda$ & \begin{minipage}[c][2cm][c]{.15\textwidth}
\centering
      \includegraphics[height=19mm]{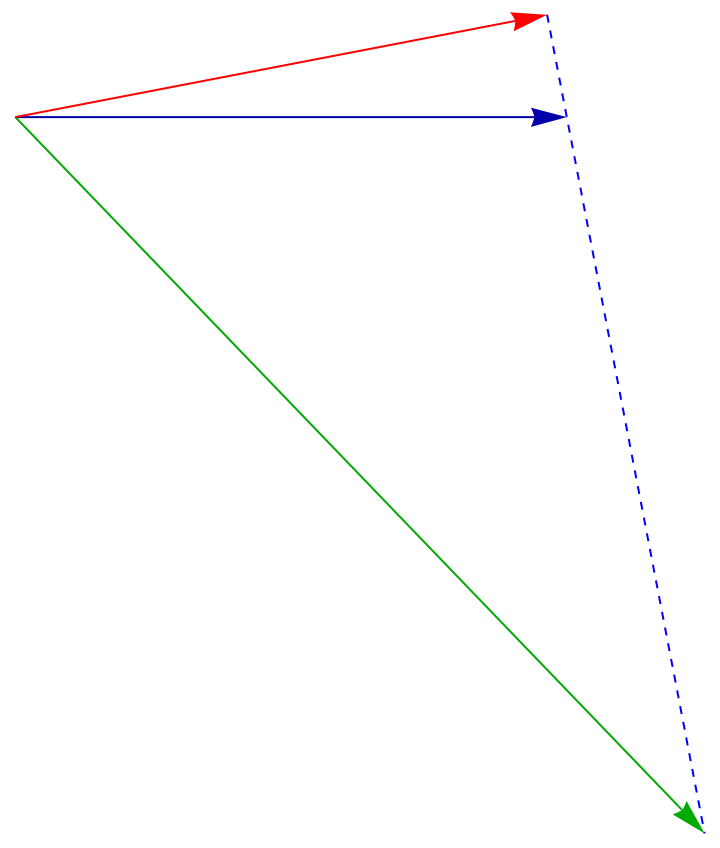}
    \end{minipage}     & N & $\vb$ &  $\frac83$   & $t^{0}$&$t^1$ &$t^{0}$  & Y & R \\[7pt]
\hline
\end{tabular}
\end{center}
\caption {Two-exponential models from universal compactifications.~The constants $A,B$ of each model are given   in terms of the 10d fluxes, cf.~Table~\ref{tab:constants}.~The vectors $\va$, $\vb$, $\vg_\perp$ are depicted in green, blue, red respectively.~The column ``$\vg_\perp$'' indicates whether or not (Y/N) the vector $\vg_\perp$ lies within the allowed range for $\vg$. 
For all models  $A>0$, and we are in the case of Fig.~\ref{fig8}.  
The column ``$\vg_*$'' indicates the value of $\vg$ at the stable critical point.~For all models  $\gamma_*^2>2$, and  the stable critical point is $P_1$.~The  models with 6d curvature must  have $\lambda<0$ for $P_1$ to be stable.~The late time scalings of $g_s$, $\exp({\sqrt{\frac{2}{3}}\varphi_1})$, $\frac{L_6(t)}{L_H}$ are listed in the respective columns.~In the column ``$g_s,\alpha'$'' we indicate whether or not  these higher-order corrections can  be tuned to be small at late times.~The column ``Sc.~S.'' indicates whether or not  scale separation can be achieved at late times;  ``R'' denotes the cases where  relative scale separation might be  possible.}
\label{tab:modelscan}
\end{table}
\renewcommand{\arraystretch}{1}

\section{Multi-exponential potential}\label{sec:multi}

\begin{figure}[H]
\begin{subfigure}{.5\textwidth}
  \centering
  \includegraphics[width=0.7\linewidth]{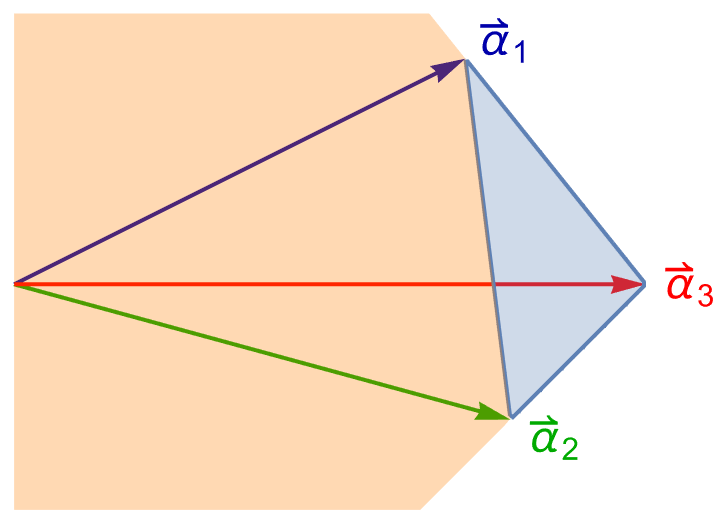}
  \caption{The case  $\Lambda_3<0$, $\Lambda_{1}, \Lambda_2>0$.}
  \label{fig:3expa}
\end{subfigure}%
\begin{subfigure}{.5\textwidth}
  \centering
  \includegraphics[width=0.69\linewidth]{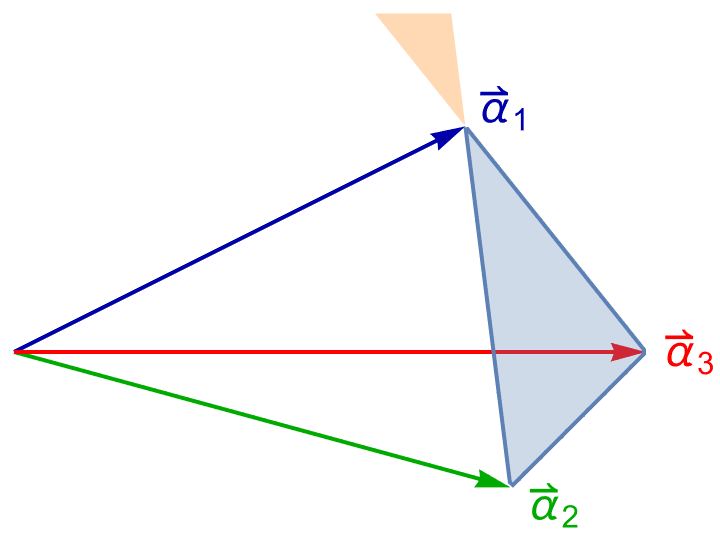}
  \caption{The case $\Lambda_1>0$, $\Lambda_2,\Lambda_3<0$.}
  \label{fig:3expc}
\end{subfigure}
\caption{Allowed region of motion for $\vg$, for a three-exponential potential.~The blue shaded region is the convex hull of the exponents; it corresponds to the allowed region for the case  $\Lambda_i>0$ for $i=1,2,3$. The allowed region in the case 
 $\Lambda_3<0$, $\Lambda_{1}, \Lambda_2>0$ ($\Lambda_1>0$, $\Lambda_2,\Lambda_3<0$) corresponds to  the  
 brown shaded region of  Figure~\ref{fig:3expa} 
 (Figure~\ref{fig:3expc}).}
\label{fig:3exp}
\end{figure}

Let us now briefly consider the case of a potential with $n$ exponential terms, function of the two-component scalar $\vv$ as before, 
\eq{
V(\varphi)=\sum_{i=1}^n\Lambda_i~\! e^{-\va_i\cdot\vv}~; ~~~V\geq0
~,}
where some of the constants $\Lambda_i$ may have  negative signs.~In the generic case, we expect 
the potential to  be dominated near the stable critical point by at most two exponentials ($\va_1$, \!$\va_2$, for the example of Figure~\ref{fig:3exp}), and our  previous results  apply.~We will therefore not enter into a detailed analysis here.~More 
than two exponentials  may  contribute asymptotically in special cases where more than two  exponents $\va_i$ lie on the same line in $\vg$-space.

Definition~\eqref{26} now gives, 
\eq{
\vg=\frac{1}{V}\sum_{i=1}^n\va_i ~\!\Lambda_i ~\! e^{-\va_i\cdot\vv}
~.}
The allowed region for $\vg$ can be determined with similar reasoning as in Section~\ref{sec:2exp}.~We have depicted in Figure~\ref{fig:3exp} the case of a three  exponential potential ($n=3$), with different possible signs for $\Lambda_i$. 
In the case $\Lambda_i>0$ for $i=1,2,3$, the allowed region for $\vg$  is the convex hull of the exponents $\va_i$ — the shaded blue region of Figure~\ref{fig:3exp}, 
\eq{\mu_1\va_1+\mu_2\va_2+(1-\mu_1-\mu_2)\va_3~;~~~\mu_1+\mu_2\leq1~;~~~\mu_{1}, \mu_2\geq0
~.}
The case $\Lambda_3<0$, $\Lambda_{1}, \Lambda_2>0$ is depicted in Figure~\ref{fig:3expa}. The allowed region for $\vg$ in this case (shaded brown region) is given by, 
\eq{\mu_1\va_1+\mu_2\va_2+(1-\mu_1-\mu_2)\va_3~;~~~\mu_1+\mu_2\geq1~;~~~\mu_{1}, \mu_2\geq0
~.}
Similarly, for $\Lambda_1>0$, $\Lambda_2,\Lambda_3<0$, the allowed region for $\vg$ is given by, 
\eq{(1-\mu_2-\mu_3)  \va_1+\mu_2\va_2+\mu_3\va_3~;~~~ \mu_{2}, \mu_3\leq0
~,}
which corresponds to  the brown shaded region of Figure~\ref{fig:3expc}.


 %

\section{Conclusions}\label{sec:conslusions}

We have studied in detail the universal cosmology models of \cite{Marconnet:2022fmx}, from the general standpoint  of a  4d gravitational theory with two minimally coupled scalars, focusing on the case of negative 3d spatial curvature.~We have 
employed a set of  phase space variables that allow for an intuitive interpretation of motion in phase space, and a particularly simple 
description of the attractor critical  point — responsible for the  late time  properties of the cosmological evolution — in terms of the convex hull of the exponents of the potential. 


Our analysis of critical points and stability is in agreement with the literature, whenever they overlap.~In particular we find that the stability of  critical points of the potential for which both exponentials contribute at late times, cf.~Table~\ref{tab:5}, requires both terms to be positive, 
as recently emphasized in \cite{VanRiet:2023cca} in the case of flat cosmologies,  in agreement with the analysis in  \cite{Hartong:2006rt}.~Furthermore, stable points with one positive and one negative exponential  are also  possible, cf.~Table~\ref{tab:6}, provided  the potential is dominated by   the positive exponential asymptotically.~Such critical points were called {\it nonproper} in \cite{Hartong:2006rt},  as  some scalars become infinite at the critical point. 
It is also straightforward to verify that the universal acceleration bound  given in   \cite[(II.10)]{Shiu:2023fhb}  for   a flat cosmology, is satisfied.\footnote{
The acceleration bound  given in   \cite[(II.10)]{Shiu:2023fhb} is saturated by the stable critical point $P_2$ of Table~\ref{tab:5}:~in this case our $\gamma_*=\gamma_{ch}$ coincides with $\hat{\gamma}_\infty$ of that reference. Moreover,   the bound is also saturated for the critical point $P_1$ of Table~\ref{tab:5}, which corresponds to an   open cosmology. On the other hand, for the case of one positive and one negative exponential,  \cite[(II.10)]{Shiu:2023fhb} provides  only a lower bound.~Indeed $\hat{\gamma}_\infty$  of  Figs.~9, 10 and 11 of \cite{Shiu:2023fhb}   corresponds to an unstable $\vg_*$, as in our Figs.~\ref{fig5a} and \ref{fig3b} respectively. }


We have seen that, in all but one model, decompactification is avoided, and both higher-order $g_s$- and   $\alpha'$-corrections 
 are suppressed at late times.~Moreover,  three of these  models exhibit time-evolution driven scale separation at late times.~The models that avoid decompactification at late times but do not exhibit scale separation,~all have potentials that contain a 6d curvature contribution that remains non-trivial at late times.~We 
 have noted that  this contribution does not need to dominate at late times for the absence of scale separation to occur, generalizing the  observations of  
 \cite{Andriot:2025cyi}, in line with the results of \cite{Shiu:2023fhb} for flat universe.~Furthermore we have discussed some mechanisms by which it might be possible to achieve relative 
 (as opposed to parametric) scale separation in these models.

 As a side note, we have constructed the fully analytic expression  of the  eternally accelerating cosmology of \cite{Andersson:2006du}.~As was noted in \cite{Marconnet:2022fmx}, 
 this cosmology can be geodesically completed in the past, and has no Big Bang singularity:~in the vicinity of the apparent singularity, the space becomes de Sitter in hyperbolic slicing.

As a general rule, the presence of multiple positive exponential terms    leads to  the  lowering of the effective exponent of the potential near the stable  critical  point.~The smallest value of  the effective exponent    
is  attained for the model of  \cite{Marconnet:2022fmx}  coming from compactification on an Einstein space with negative curvature and non-vanishing Romans mass, and is 
 smaller than the threshold   below which the attractor point  $P_1$  is a stable node (but above the swampland bound).~Multiple exponentials    thus provide a mechanism that allows to lower the 
  curvature energy density near the critical point, potentially improving the compatibility of these models  with observation.

Our analysis has mostly 
focused on two-exponential potentials, which is  the case that  captures all  the essential physics at late times.~Although  we do not expect the multi-exponential  case to  alter the physics asymptotically, it is a different story when it comes to detailed universe histories in the bulk of phase space. Indeed the latter would become important if one wishes to 
 generalize the present analysis to a multi-exponential quintessence scenario with the inclusion of a  general barotropic fluid, with the aim of modeling radiation and matter throughout the history of the universe  — 
 as was done in \cite{Andriot:2024jsh} for the case of a single exponential.

\section*{Acknowledgment}

We would like to thank George Tringas for collaboration on inflationary aspects of the models considered here, and  David Andriot for useful discussions on 
cosmological scale separation.

\bibliographystyle{JHEP}

\providecommand{\href}[2]{#2}\begingroup\raggedright\endgroup

\end{document}